\documentclass[10pt]{article}

\usepackage{amsmath}
\usepackage{amssymb}

\usepackage{graphicx}

\usepackage{cite}

\usepackage{color} 

\usepackage[labelfont=bf,labelsep=period,justification=raggedright]{caption}

\makeatletter
\renewcommand{\@biblabel}[1]{\quad#1.}
\makeatother

\date{}

\begin{document}

% Title must be 150 characters or less
\begin{flushleft}
{\Large
\textbf{Neurogenesis Drives Stimulus Decorrelation in a Model of the Olfactory Bulb}
}
% Insert Author names, affiliations and corresponding author email.
\\
Siu-Fai Chow$^{1}$, 
Stuart D. Wick$^{3}$, 
Hermann Riecke$^{1,2,\ast}$
\\
\bf{1} Engineering Sciences and Applied Mathematics, Northwestern University, Evanston, IL 60208, USA
\\
\bf{2} Northwestern Institute on Complex Systems, Northwestern University, Evanston, IL 60208, USA
\\
\bf{3} Department of Physics, North Central College, Naperville, IL 60540, USA
\\
$\ast$ E-mail: Corresponding h-riecke@northwestern.edu
\end{flushleft}

% Please keep the abstract between 250 and 300 words
\section*{Abstract}

The reshaping and decorrelation of similar activity patterns by neuronal
networks can enhance their discriminability, storage, and retrieval.
How can such networks learn to decorrelate new complex patterns, as
they arise in the olfactory system? Using a computational network
model for the dominant neural populations of the olfactory bulb we
show that fundamental aspects of the adult neurogenesis observed in
the olfactory bulb -- the persistent addition of new inhibitory granule
cells to the network, their activity-dependent survival, and the reciprocal
character of their synapses with the principal mitral cells -- are
sufficient to restructure the network and to alter its encoding of
odor stimuli adaptively so as to reduce the correlations between the
bulbar representations of similar stimuli. The decorrelation is quite
robust with respect to various types of perturbations of the reciprocity. The model parsimoniously captures the experimentally observed role
of neurogenesis in perceptual learning and the enhanced response of
young granule cells to novel stimuli. Moreover, it makes specific
predictions for the type of odor enrichment that should be effective
in enhancing the ability of animals to discriminate similar odor mixtures. 

% Please keep the Author Summary between 150 and 200 words
% Use first person. PLoS ONE authors please skip this step. 
% Author Summary not valid for PLoS ONE submissions.   
\section*{Author Summary}

The olfactory bulb is one of only two brain regions in which new neurons
are added persistently in substantial numbers even in adult animals.
This leads to an ongoing turnover of interneurons, in particular of
the inhibitory granule cells, which constitute the largest cell population
of the olfactory bulb. The function of this adult neurogenesis in
olfactory processing is only poorly understood. Experiments show that
it contributes to perceptual learning. We present a basic computational
model that is built on fundamental aspects of the granule cells and
their connections with the excitatory mitral cells, which convey the
olfactory information to higher brain areas. We show that neurogenesis
can reshape the network connectivity in response to olfactory input
so as to reduce the correlations between the bulbar representations
of even highly similar stimuli. The neurogenetic adaptation of the stimulus representations provides a natural explanation of the perceptual learning and the different response of young and old granule cells to novel odors that have been observed in experiments. The model makes experimentally testable predictions for training protocols that enhance the discriminability of odor mixtures. 

\section*{Introduction}

Contrast enhancement and decorrelation are common steps in information
processing. They can reshape neuronal activity patterns so as to enhance
down-stream processing like pattern discrimination, storage, and retrieval.
The activity patterns can be complex and new patterns may become relevant
due to changes in the environment or in the life circumstances of
the animal. How can networks adapt to such demands, as they arise,
for instance, in the olfactory system? What are neural substrates
that would allow the necessary network restructuring? 

In the olfactory system initial sensory processing is performed in
the olfactory bulb. Its inputs consist of activation patterns of its
100-1,000 glomeruli, each of which can be considered as an individual
input channel representing a specific olfactory receptive field. The
bulbar network reshapes the patterns representing odor stimuli and
typically reduces the correlation between output patterns representing
similar odors as compared to the respective input patterns \cite{FrLa01,FrHa04,WiJu10}.
It does so despite the fact that even simple odors evoke complex activation
patterns due to the fractured representation of the high-dimensional
odor space on the two-dimensional glomerular surface \cite{ClSe06}.
Unlike spatial contrast enhancement in the retina \cite{Ku53}, this
decorrelation can therefore not arise from local lateral inhibition
that is confined to neighboring glomeruli \cite{FaSo08,WiJu10}. What
types of network connectivities can then underlie the enhancement
of small, but significant differences in the representation of similar
odors? 

Previously, a number of different decorrelation mechanisms have been
proposed, each of which exploiting a different aspect of the nonlinear
dynamics of the bulbar network. The network connectivities were taken
to be fixed, either without any lateral inhibition \cite{ClSe06},
with all-to-all inhibition \cite{ArKa08}, or with sparse random connections
across large portions of the bulb \cite{WiJu10}. These networks were
shown to reduce quite effectively the correlation between the representations
of moderately similar stimuli. 

A different perspective is suggested by two distinctive features of
the olfactory system: i) many odors do not have an intrinsic meaning
to the animal and their significance is likely to be learned by experience
\cite{WiSt06a,BrKe06,MaWe10}; ii) the bulbar network structure is
not static but undergoes persistent turn-over due to neurogenesis
and apoptosis even in adult animals \cite{WhGr09,OrMu07}. 

So far, the specific role of adult neurogenesis for olfactory processing is only poorly understood \cite{LlAl06,LazLle11}.
It is known that environmental changes like sensory deprivation \cite{PeAl02,MaJo03,YaMo05,MaSa06a}
and odor enrichment \cite{RoGh02,MaMi05,AlOr08}, associative learning
\cite{DoDr03,MoGh08,KeSu10,SulDid11}, and life circumstances like
mating \cite{MaEn07} and pregnancy \cite{ShGr03} affect anatomical
and functional aspects of the olfactory bulb. Moreover, genetic \cite{KiKi07,ImSa08},
pharmacological \cite{BrLe09,MoLe09,SuMa10}, and radiational manipulations
\cite{LaMo09,VaMu09} have identified the significance of neurogenesis
in these modifications. 

Here we ask whether the neuronal turnover associated with adult
neurogenesis can provide a neural substrate for the adaptation of the
network to the decorrelation of different relevant stimuli that may be
highly similar. Such a contribution of neurogenesis to pattern
separation has been proposed for the olfactory bulb as well as the
dentate gyrus \cite{SahHen11}. We use a minimal computational network
model of neurogenesis in the olfactory bulb that incorporates the
persistent addition of new inhibitory interneurons (granule cells)
into the olfactory bulb \cite{NiMo07}, their connection with the
principal mitral cells via \emph{reciprocal} synapses through which
the mitral cells excite the granule cells and the granule cells
inhibit the mitral cells \cite{Sh04}, and the activity-dependent
apoptosis of the granule cells
\cite{PeAl02,WiCo02,MaDi08,MoLi09,SuMa10,LiSi10}. Using stimulus
ensembles based on glomerular excitation patterns observed in rat
\cite{JoLe07} we find that the networks learn to decorrelate even very
similar stimuli. This results largely from the surviving granule cells
detecting strongly co-active mitral cells and providing lateral
inhibition between them. Our modeling gives a natural interpretation
of recent experiments on the role of neurogenesis in the perceptual
learning of a non-associative odor discrimination task \cite{MoLi09}
and the detection of novel odors \cite{MaMi05}. Our computational
model predicts that learning to decorrelate highly similar mixtures
comprised of dissimilar components requires the exposure to a mixture
of the components rather than the individual components themselves.
This can be tested in behavioral experiments using suitable enrichment
protocols \cite{MaSt06,MaSt06a,MaSt06b,MoLi09}.

% Results and Discussion can be combined.
\section*{Results}

\subsection*{Activity-Dependence of Survival Drives Decorrelation}
\begin{figure}[!ht]
\begin{center}
\includegraphics[width=14cm]{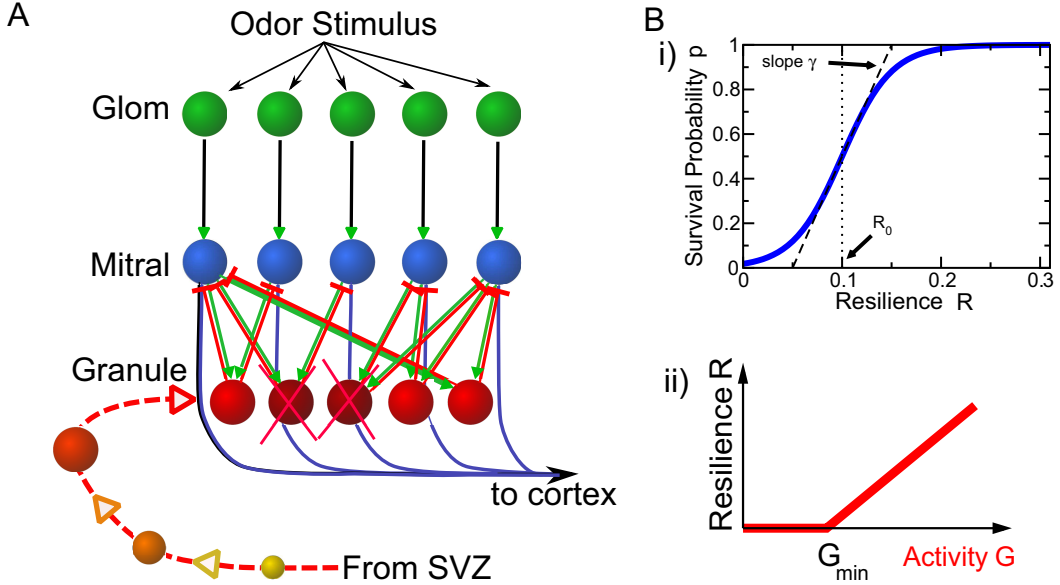}
\end{center}
\caption{
{\bf Main components of the model.} \textbf{A) }Sketch of the recurrent bulbar network model with neurogenesis.
Odor stimuli evoke glomerular activation patterns (green). The glomeruli
drive mitral cells (blue), which relay the information to cortex.
In addition, they excite granule cells (red), which through their
reciprocal synapses provide self-inhibition and lateral inhibition
to the mitral cells. New granule cells migrate persistently from the
subventricular zone to the olfactory bulb and are incorporated into
the network (yellow to orange). They are removed if their activity
is too low (dark red). \textbf{B) }The survival probability of granule
cells depends sigmoidally on their resilience (i), which is a threshold-linear
function of their activity (ii), summed over the stimulus ensemble.
}
\label{fig:network_sketch}
\end{figure}

In our computational model we consider the recurrent network formed
by principal mitral cells and inhibitory granule cells. We focus on
the adaptive restructuring of the network connectivity in response
to a stimulus ensemble and model the individual neurons in a minimal
fashion using linear firing-rate dynamics (cf. \textit{METHODS, Discrete Adaptive Network Model}). Focusing on the evolution of the network structure we ignore transients in the evolution of the neuronal activities and consider only their steady states in response to any given odor stimulus. The network
is persistently rewired by adding in each time step of the network
evolution randomly connected new granule cells and removing granule
cells that are not sufficiently active during the steady state reached
in response to odor stimulation (Fig.\ref{fig:network_sketch}).
Specifically, the survival probability of a granule cell depends in
a sigmoidal fashion on its `resilience' $R$, which we introduce as
its thresholded activity summed over the stimulus ensemble.

\begin{figure}[!h]
\begin{center}
\includegraphics[width=14cm]{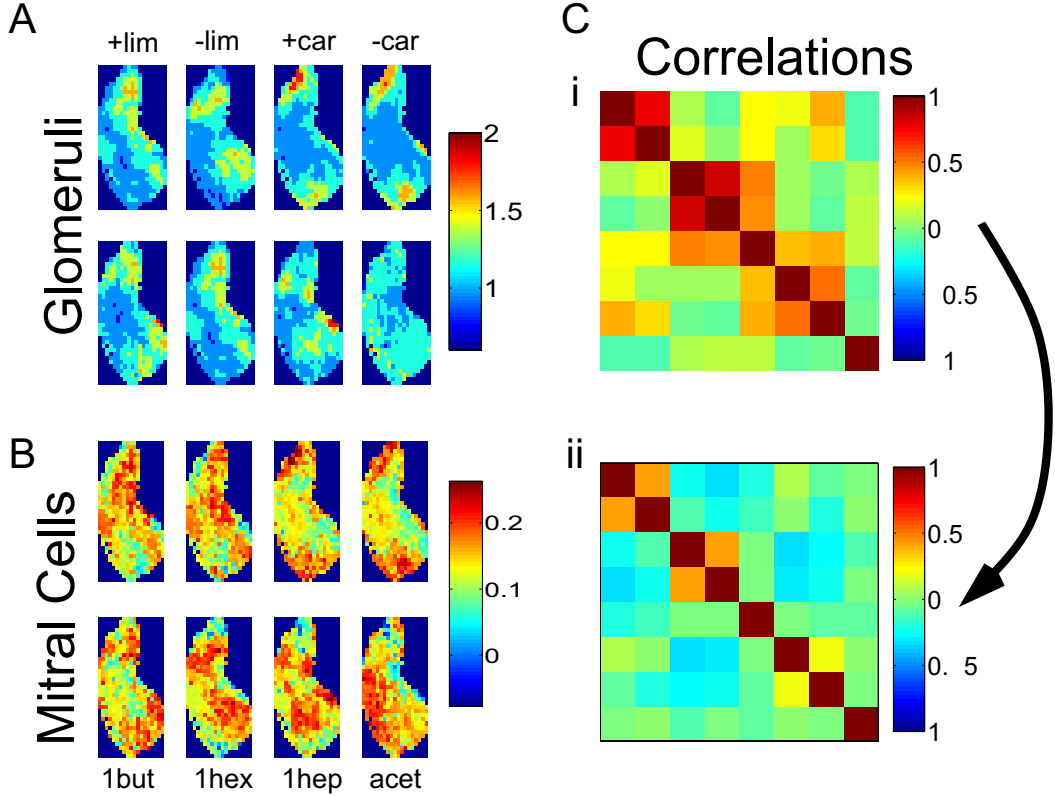}
\end{center}
\caption{
{\bf Decorrelation of natural stimuli.} \textbf{A)} Glomerular activation
patterns in rat for the odorants $\pm$-limonene, $\pm$-carvone,
1-butanol, 1-hexanol, 1-heptanol, and\textbf{ }acetic acid \cite{JoLe07}.
\textbf{B)} Mitral cell activity patterns of a network trained on
all eight stimuli. \textbf{C) }Correlation matrix of the input patterns
(i) and of the mitral cell output patterns (ii). The stimuli are ordered
as in A and B. Parameters: $\gamma=20$, $R_{0}=0.1$, $G_{min}=1.2$,
$N_{m}=424$, $n_{connect}=8$, $\beta=33$, $w=0.005$, $M_{sp}=1$.
}
\label{fig:mouse_stimuli_decorrelation}
\end{figure}

In most of the computations we use input patterns that are based on
a set of experimentally obtained glomerular activity patterns in rat
\cite{JoLe07} corresponding to the odorants $\pm$-limonene, $\pm$-carvone,
1-butanol, 1-hexanol, 1-heptanol, and\textbf{ }acetic acid (Fig.\ref{fig:mouse_stimuli_decorrelation}A).
They drive 424 mitral cells, which in turn excite about 10,000 granule
cells. Due to the reciprocal character of these synapses each granule
cell provides self-inhibition to each of the eight mitral cells that
drive it as well as lateral inhibition between them (Fig.\ref{fig:network_sketch}A).
All synaptic strengths are taken to be  fixed. Unless noted otherwise,
all excitatory and all inhibitory synapses have equal strengths, respectively.

The network that eventually emerges as a statistically steady state
from the persistent rewiring substantially reshapes the representation
of the stimuli (Fig.\ref{fig:mouse_stimuli_decorrelation}B). In particular,
the mitral cell activation patterns, which represent the output of
the olfactory bulb, differ from each other significantly more than
the glomerular input patterns. To quantify this reduction in similarity
we use the Pearson correlation $r_{\alpha\beta}$ of the patterns
associated with stimuli $\alpha$ and $\beta$ (cf. Eq.(\ref{eq:Pearson})),
as has been done in previous, experimental studies \cite{FrLa01,FrHa04,WiJu10}.
Thus, the network achieves a substantial decorrelation of the stimulus
representations (Fig.\ref{fig:mouse_stimuli_decorrelation}Ci,ii).
This is the case for the highly similar $\pm$-limonene- and $\pm$-carvone-pairs
as well as the less correlated, remaining stimuli of the odor ensemble.
Moreover, through the enhanced inhibition of mitral cells that are
strongly driven in this stimulus ensemble and the spontaneous activity
of mitral cells that receive very little or no input \cite{RiKo06a,WiJu10}
the network reshapes the quite focal input patterns into output patterns
in which the activity is more broadly distributed over the whole
network (Fig.\ref{fig:mouse_stimuli_decorrelation}B). Such a reduction
of the focality of the output patterns has been observed for mitral
cell activity in zebrafish \cite{WiJu10}. Particularly for stimuli
that predominantly overlap in these focal areas such a reshaping of
the pattern can reduce the correlation significantly. 

\begin{figure}[!h]
\begin{center}
\includegraphics[width=14cm]{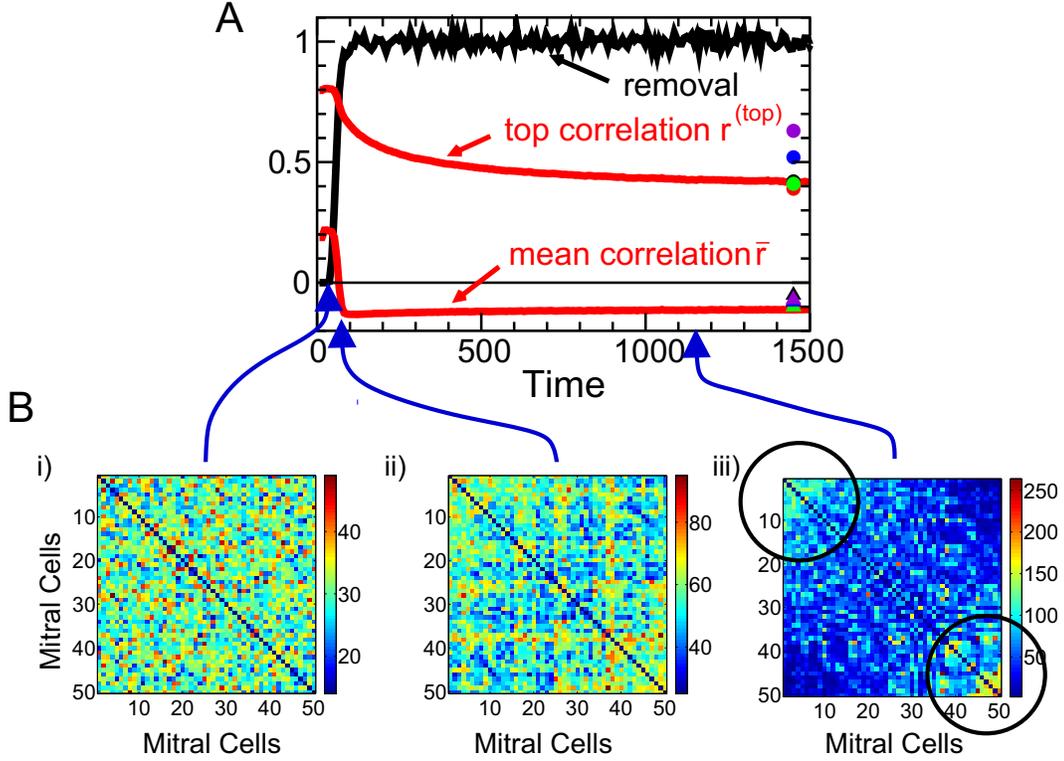}
\end{center}
\caption{
{\bf Decorrelation and connectivity.} Evolution of the pattern correlation
and rate of granule cell removal (scaled by their influx) (\textbf{A}),
and the effective connectivity matrix $\boldsymbol{\mathcal{W}}$
between pairs of mitral cells (cf. Eq.(\ref{eq:effective_connectivity}))
(\textbf{B}). Initially ($t<40$) almost all granule cells survive,
generating a random connectivity that does not decorrelate the stimuli
(\textbf{Bi}). By $t=100$ the selective removal of weakly active
granule cells leads to a structured connectivity (\textbf{Bii}) that
reduces the mean correlation $\bar{r}$. The highly similar stimuli
$\pm$-limonene and $\pm$-carvone are only decorrelated by strong
inhibition between highly co-active mitral cells (marked by black
circles), which emerges in the final steady state (\textbf{Biii}).
Parameters for the simulation in \textbf{A} as in Fig.\ref{fig:mouse_stimuli_decorrelation}.
The correlations have been averaged over 16 runs. The symbols at\textbf{
}$t=145$0 denote output correlations for different slopes of the
survival curve, $\gamma=20,10,5,$2.5,1 (cf. Fig.\ref{fig:network_sketch}Bi).
For visual clarity the connectivities are shown in \textbf{B} for
a reduced network of 50 instead of 424 mitral cells (for parameters
see Fig.S1B in Text S1). In the connectivity
matrices the diagonal elements have been divided by 10. 
}
\label{fig:network_evolution}
\end{figure}

Insight into the mechanisms underlying the decorrelation by the network
is gained by following the evolution of the connectivity and the associated
decorrelation performance as the network builds up from a network
without any granule cells (Fig.\ref{fig:network_evolution}). The
early stages of this evolution are not meant to mimic the peri-natal
development of the bulb, which is controlled by mechanisms other than
those included in this model. To visualize the network connectivities
the stimuli are down-sampled to 50\textbf{ }channels (cf. \textit{METHODS, Natural Stimuli})
and the two-dimensional activation patterns are re-arranged into one-dimensional
vectors in which the mitral cells that are strongly activated during
the $\pm$-limonene presentation are located at the beginning of the
vector and those that dominate during the $\pm$-carvone presentation
at the end. Because the overlap between the activation patterns of
these two pairs of enantiomers is small there are only few mitral
cells that receive significant input for both types of stimuli. They
end up towards the middle of the activity vector. For visual clarity
the diagonal elements of the connectivity matrices are divided by
10.

During the initial phase $0<t\lesssim40$ the granule cell population
is small and provides only little inhibition to the mitral cells.
Their activities and with them the activities of the granule cells
are therefore high and none of the granule cells are removed (Fig.\ref{fig:network_evolution}A).
Since the granule cells establish random connections with the mitral
cells the resulting effective connectivity between the mitral cells
is essentially random (Fig.\ref{fig:network_evolution}Bi) and the
activity patterns are only reduced in amplitude without any qualitative
changes; the correlations remain high. As the mitral cell activities
decrease, some granule cells fall in their activity and resilience
below the soft survival threshold $R_{0}$  (cf. Fig.\ref{fig:network_sketch}Bi)
and their survival probability drops drastically ($t\gtrsim40)$.
This apoptosis is selective, resulting in a structured connectivity,
in which more highly active mitral cells receive stronger inhibition
(Fig.\ref{fig:network_evolution}Bii), and a reduction of the mean
pattern correlation. The correlation between the highly similar stimuli
is, however, still high. In the third phase of the network evolution
the size of the granule cell population remains constant, but the
connectivity evolves slowly towards establishing strong effective
mutual inhibition between mitral cells that are highly co-active during
$\pm$-limonene or $\pm$-carvone presentations (marked by circles
in Fig.\ref{fig:network_evolution}Biii). In parallel, the correlation
$r^{(top)}$ of these highly similar enantiomers is strongly reduced. 

\begin{figure}[!h]
\begin{center}
\includegraphics[width=14cm]{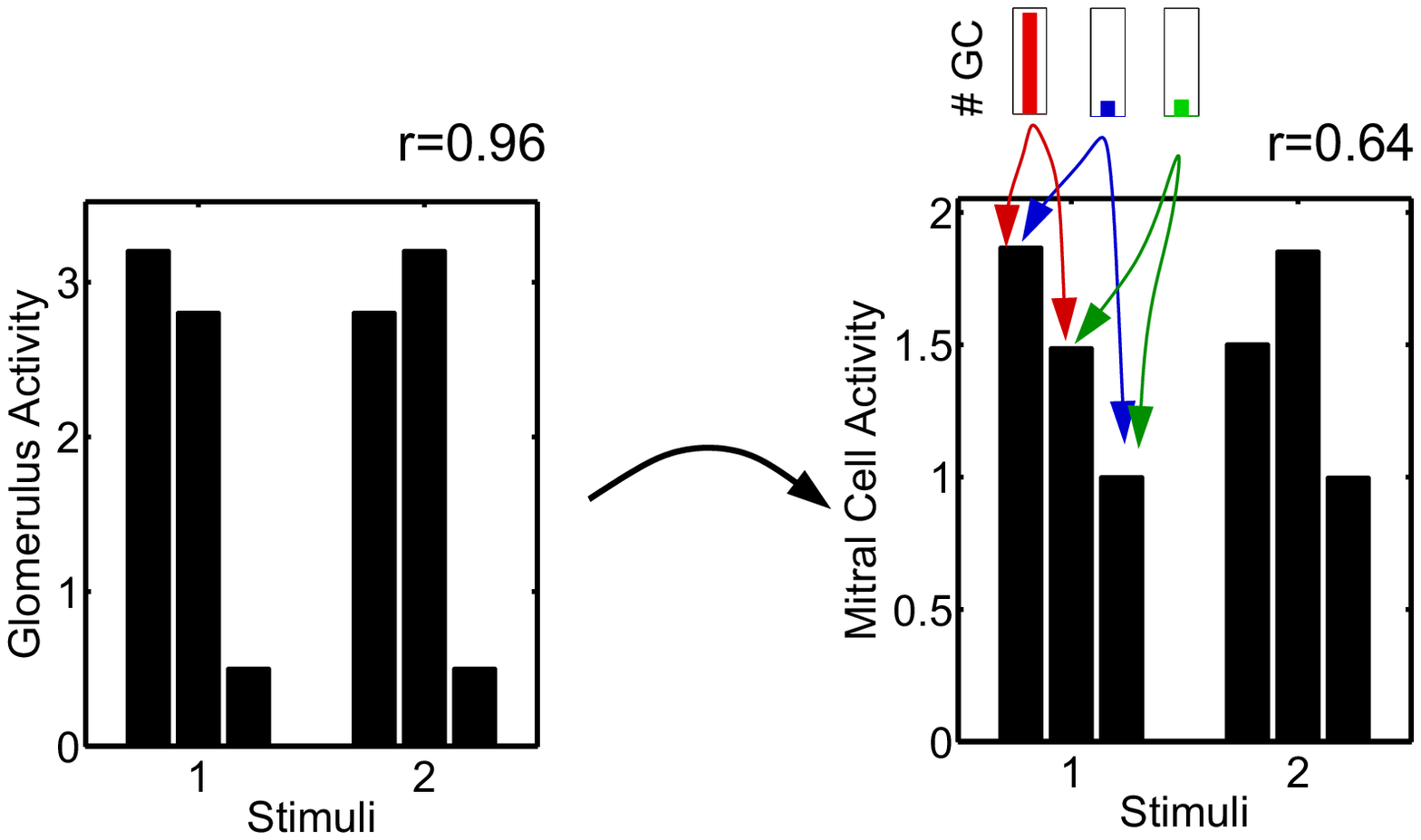}
\end{center}
\caption{
{\bf Decorrelation by inhibition of strongly co-active mitral cells.} Since
mitral cells 1 and 2 are strongly driven in both stimuli the population
of granule cells (GC) connected to these mitral cells (red) is much
larger than the other two populations (blue, green). The associated
inhibition strongly suppresses the activities of mitral cells 1 and
2, but not of mitral cell 3, which reduces the correlation of the
patterns from $r=0.96$ to $r=0.64$. The mitral cells have a spontaneous
activity $M_{sp}=1$.
}
\label{fig:decorrelation_cartoon}
\end{figure}

The effectiveness of the inhibition of highly co-active mitral cells
in decorrelating activity patterns is illustrated using a very simple
example with stimuli exciting only three mitral cells (the relevant
two stimuli are shown in Fig.\ref{fig:decorrelation_cartoon}). Like
the highly similar olfactory stimuli in Fig.\ref{fig:mouse_stimuli_decorrelation},
these stimuli overlap in strongly co-active glomeruli. This allows
the population of granule cells that connect to the mitral cells driven
by these glomeruli to be much larger than the other two populations.
The reciprocity of the synapses implies that these mitral cells receive
substantially stronger inhibition than the other mitral cell. The
resulting reduction in amplitude reduces also the correlation between
the two mitral cell activity patterns. In Fig.\ref{fig:network_evolution}Biii
the corresponding enhanced connectivity between mitral cells that
are highly co-active during $\pm$-limonene (or $\pm$-carvone) stimulation
is marked by black circles.

\subsection*{Threshold Promotes  Lateral Inhibition Based on Co-Activity}

What determines the performance of the networks arising from the persistent
turn-over? The granule-cell survival is controlled by two thresholds:
i) for each stimulus for which the granule cell activity surpasses
the resilience threshold $G_{min}$ its resilience $R$ increases
(cf. Eq.(\ref{eq:resilience})) and ii) the resilience accumulated
across all stimuli of the ensemble has to be above the soft survival
threshold $R_{0}$ in order for the granule cell to have a significant
survival probability (cf. Eq.(\ref{eq:survival})). The survival threshold
$R_{0}$ controls in particular the total number of granule cells
and with it the overall level of inhibition. In general, the overall
correlation of the outputs decreases with increasing inhibition (data
not shown) at the expense of the output amplitudes. In our comparisons
we adjust therefore $R_{0}$ to keep the mean output amplitudes fixed. 

\begin{figure}[!h]
\begin{center}
\includegraphics[width=14cm]{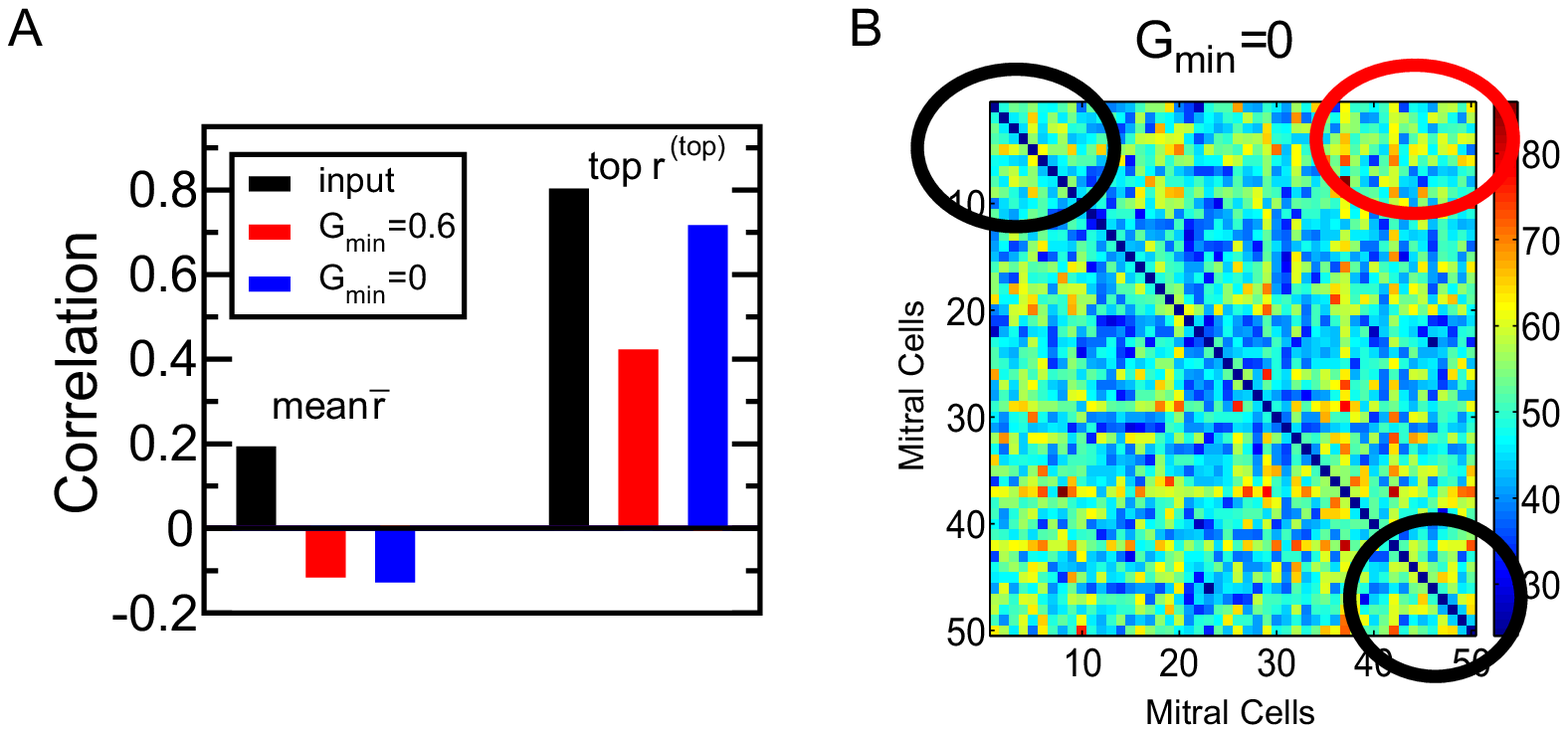}
\end{center}
\caption{
{\bf Resilience threshold $G_{min}$ reduces interference and enhances
decorrelation of highly correlated stimuli.} \textbf{A) }For $G_{min}=0$
the networks achieve the same level of overall decorrelation as networks
with suitable $G_{min}>0$, but they decorrelate the representations
of highly similar stimuli very poorly. Parameters: $\gamma=20$, $R_{0}=4$,
$G_{min}=0$, $N_{m}=424$, $n_{connect}=8$, $\beta=33$, $w=0.0057$,
$M_{sp}=1$. For $G_{min}=0.6$ parameters as in Fig.\ref{fig:mouse_stimuli_decorrelation}.
\textbf{B)} Effective connectivity matrix $\mathcal{W}$  for $50$
mitral cells with diagonal elements divided by 10.\textbf{ }For $G_{min}=0$
the interfering connections between mitral cells that are active for
$\pm$-limonene \emph{or} for $\pm$-carvone (red ellipse) are as
strong as those between co-active cells (black ellipses); cf. panel
bottom right on Fig.\ref{fig:network_evolution}. Parameters: $\gamma=20$,
$R_{0}=2.25$, $G_{min}=0$, $N_{m}=50$, $n_{connect}=4$, $\beta=16$,
$w=0.0023$, $M_{sp}=1$. 
}
\label{fig:threshold}
\end{figure}

A more subtle and interesting role is played by the resilience threshold
$G_{min}$. For $G_{min}=0$ the network achieves an overall decorrelation
that is quite comparable to that of the network of Fig.\ref{fig:mouse_stimuli_decorrelation}
with $G_{min}=0.6$; the highly similar stimuli $\pm$-limonene and
$\pm$-carvone, however, are only very poorly decorrelated (Fig.\ref{fig:threshold}A).
The origin of this poor performance is apparent in the effective connectivity
obtained with $G_{min}=0$ (Fig.\ref{fig:threshold}B). A comparison
with the connectivity arising for $G_{min}=0.6$ (Fig.\ref{fig:network_evolution})
reveals that the  connections among the mitral cells that are co-active
in response to $\pm$-limonene (or $\pm$-carvone) stimulation (black
circles) are not stronger than among mitral cells that are not co-active
(red circle).  As had been observed in Fig.\ref{fig:network_evolution},
it is the connections among co-active mitral cells, however, that
are essential for decorrelating these stimulus representations.  

\begin{figure}[!h]
\begin{center}
\includegraphics[width=14cm]{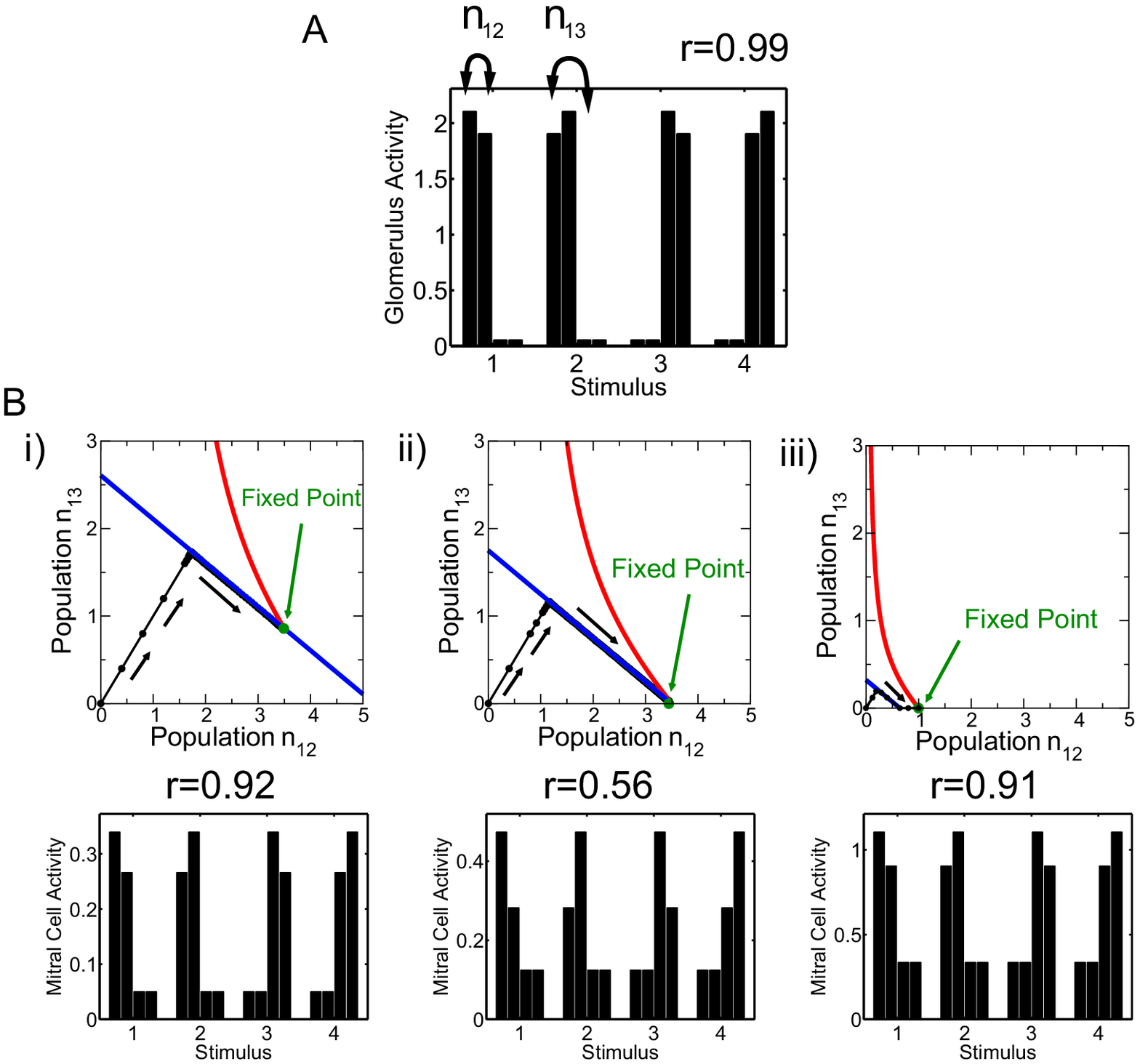}
\end{center}
\caption{
{\bf Interference and optimal resilience threshold $G_{min}$.} \textbf{A)}
Two pairs of symmetrically related stimuli comprised of four glomeruli
each (cf. eqs.(\ref{eq:4_stimuli_1},\ref{eq:4_stimuli_2}). The granule
cells are described by the populations $n_{12}=n_{34}$ and $n_{13}=n_{23}=n_{14}=n_{24}$.
Stimulus pairs $\mathbf{S}^{(1,2)}$ and $\mathbf{S}^{(3,4)}$ are
highly correlated ($r=0.99$). \textbf{B) }Top panels: Phase plane
with nullclines $R_{12}=R_{0}$ (red) and $R_{13}=R_{0}$ (blue) and
the trajectory $(n_{12}(t),n_{13}(t))$ (black symbols) starting from
$n_{12}=0=n_{13}$ and ending up on the fixed point. The network evolution
is indicated by black arrows. Bottom panels: mitral cell activity
patterns. i) $G_{min}=0.1$. Interference ($n_{13}>0$) strongly suppresses
the weakly driven mitral cells. High correlation ($r=0.92$).\textbf{
}ii) $G_{min}=G_{min}^{(opt)}=0.25$. No interference ($n_{13}=0$),
but strong inhibition among the highly co-active mitral cells through
large population $n_{12}$. Low correlation ($r=0.56$). iii) $G_{min}=1.5$.
The inhibition of co-active mitral cells is weak. High correlation
($r=0.91$). Other parameters: $M_{sp}=1$, $S=2$, $s=0.5$, $R_{0}=1$,
$\beta=0.001$, $\gamma=500$. 
}
\label{fig:Phase-plane-analysis}
\end{figure}

How does the threshold $G_{min}$ provide a co-activity detector?
Why do the connections among mitral cells that are not co-active interfere
with the decorrelation? The function of the threshold can be illustrated
with a minimal set of two pairs of strongly correlated stimuli \textbf{$\mathbf{S}^{(\alpha)}$}
activating four glomeruli, $\mathbf{S}^{(1,2)}=\left(S\pm s,S\mp s,0,0\right)$
and $\mathbf{S}^{(3,4)}=\left(0,0,S\pm s,S\mp s\right)$ with $s\ll S$
(Fig.\ref{fig:Phase-plane-analysis}A). Stimuli $\mathbf{S}^{(1,2)}$
and $\mathbf{S}^{(3,4)}$ may be viewed as caricatures of the limonene
and carvone enantiomers, respectively. The granule cells in population
$n_{12}$ inhibit the mitral cells that are co-active in stimuli $\mathbf{S}^{(1,2)}$
(cf. Fig.\ref{fig:Phase-plane-analysis}A) and are therefore needed
for decorrelation. The granule cells in population $n_{13}$, however,
are connected to mitral cells that are not co-active in any of the
stimuli; they may interfere with the performance of the network. The
resilience $R_{12}$ of the granule cells in population $n_{12}$
is comprised of two large contributions due to the strong inputs in
stimuli $\mathbf{S}^{(1)}$ and $\mathbf{S}^{(2)}$ and two small
contributions from stimuli $\mathbf{S}^{(3)}$ and $\mathbf{S}^{(4)}$,
while the resilience $R_{13}$ of the cells in population $n_{13}$
is determined by 4 intermediate contributions. In our model (\ref{eq:M_eq},\ref{eq:G_eq})
for the neuronal dynamics the granule cell activities are linear in
the mitral cell activities. For $s\ll S$ the activity $G_{13}$ of
the interfering population $n_{13}$ is almost the same for all four
stimuli and is  close to the average of the activity $G_{12}$ across
the four stimuli. As a result, the rectifier, which makes the resilience
function (\ref{eq:resilience}) concave, renders the granule cells
that establish interfering connections less resilient than the granule
cells connecting co-active mitral cells, $R_{13}<R_{12}$. This suppresses
the interfering population $n_{13}$ relative to $n_{12}$, as is
apparent in a comparison of Fig.\ref{fig:network_evolution}Biii and
Fig.\ref{fig:threshold}B.

Within the framework of the population formulation eqs.(\ref{eq:population_model},\ref{eq:population_MC},\ref{eq:population_GC})
the simplicity of the minimal stimulus set of Fig.\ref{fig:Phase-plane-analysis}A
allows a detailed analysis of the role of the threshold in the balance
between the suppression of interfering connections and a reduction
of the beneficial inhibition of co-active mitral cells. Due to the
symmetry of the stimulus ensemble only two granule-cell populations
have to be analyzed, $n_{12}$ and $n_{13}$.  Their dynamics  can
be understood using a phase-plane analysis. For steep survival curves
$p(R)$ the nullclines of $n_{1j}$, which are defined by $\frac{dn_{1j}}{dt}=0$,
are very well approximated by $R_{1j}=R_{0}$ (cf. Fig.\ref{fig:network_sketch}Bi).
Starting from $n_{1j}=0$, both population sizes increase linearly
in time until they reach one of the two nullclines. Then the system
follows slowly that nullcline until a fixed point is reached. This
can be the intersection of the two nullclines (Fig.\ref{fig:Phase-plane-analysis}Bi).
In addition, since $n_{13}$ cannot become negative, an intersection
of the nullcline $R_{12}=R_{0}$ with the axis $n_{13}=0$ also represents
a fixed point if at that point $R_{13}<R_{0}$ (Fig.\ref{fig:Phase-plane-analysis}Biii)
and similarly with the roles of $n_{12}$ and $n_{13}$ interchanged.

A straightforward expansion shows that for highly similar stimuli,
$s\ll S$, the correlation between the two output patterns $\mathbf{M}^{(1,2)}$
is given by\begin{equation}
r_{12}=1-4\,\frac{\left(1+2n_{12}\right)^{2}}{\left(1+2n_{13}\right)^{2}}\frac{s^{2}}{S^{2}}.\label{eq:correlation_expansion}\end{equation}
Thus, as expected, the correlation decreases with increasing reciprocal
inhibition $n_{12}$ of co-active mitral cells and increases with
increasing strength of the interfering connections $n_{13}$. As discussed
above, the relation between these two populations can be controlled
using the threshold $G_{min}$.  For fixed resilience threshold $R_{0}$
the correlation is minimized for (cf. Eqs.(\ref{eq:optimal_Gmin-1},\ref{eq:fixed_point_population}))\textbf{
\begin{equation}
G_{min}=G_{min}^{(opt)}\equiv\frac{1}{2}\frac{R_{0}M_{sp}}{S}.\label{eq:Gmin_opt}\end{equation}
}This is the smallest value of $G_{min}$ for which the interfering
connections vanish, $n_{13}=0$. Thus,  it maximizes the inhibition
between co-active mitral cells without inducing interference. This
leads to optimal decorrelation, as is also apparent in the output
activity patterns in the bottom panels of Fig.\ref{fig:Phase-plane-analysis}B.

{Thus, the threshold $G_{min}$ in the resilience suppresses
interfering connections between mitral cells that are not co-active
and promotes a connectivity that is based on co-activity. To provide a
context of the performance of this co-activity based connectivity we
compare the decorrelation achieved by the resulting networks with that
obtained by a number of other types of adaptive networks. In some of
them the inhibition is also based on co-activity, in others on
distance, correlation, or covariance (see Text S1 with Figs.S1,S2,S3
therein). We find that the networks whose adaptation mechanism is
based on some form of co-activity of mitral cells or glomeruli are
able to decorrelate representations of highly similar stimuli and
achieve a reduction of the overall correlations without and with
significant spontaneous mitral cell activity. Among these networks are
networks motivated by an earlier model for neurogenesis \cite{CePe01}
as well as networks that aim to orthogonalize the stimulus
representations by orthogonalizing (and normalizing) the activity
vectors of pairs of mitral cells \cite{WiWi10}. Alternatively, the
connectivities can also be based on the correlations or covariances of
the inputs. For instance, a correlation-based connectivity was found
to capture the outputs of the bee antennal lobe, which is the insect
homolog of the olfactory bulb, better than random or local
connectivities \cite{LiSa05}. We find that correlation- and
covariance-based recurrent networks do not decorrelate stimulus
representations very well. In various situations they even tend to
increase rather than decrease the correlations. This reflects, in
part, the fact that they are not sensitive to the spontaneous activity
of the mitral cells.

\subsection*{Imperfect Reciprocity of Synapses is Sufficient}

Anatomically, the dendrodendritic synapses between mitral cells and
granule cells are found to be predominantly reciprocal,  i.e. each
granule cell has inhibitory connections only to those mitral cells
from which it receives excitatory connections \cite{Sh04}. In combination
with the threshold $G_{min}$ this establishes effectively inhibitory
lateral connections selectively between highly co-active mitral cells
 and allows the networks  to decorrelate their highly correlated inputs. 

As implemented in our model so far, the reciprocal synapses not only
provide an anatomical connection between co-active mitral cells but
due to the homogeneity of the inhibitory synaptic weights they also
induce a symmetric connectivity matrix and the amount of self-inhibition
that a given mitral cell experiences is directly related to the amount
of lateral inhibition it provides to other mitral cells. What roles
do these different aspects play in the decorrelation?

To test the importance of the correct anatomical connections  we 
redirect  a fraction of the inhibitory connections of each granule
cell to  randomly chosen mitral cells instead of the mitral cells
that drive that granule cell. As expected, as the fraction of such
non-reciprocal synapses increases the  correlations increase as well.
Without any reciprocal synapses the network does not decorrelate the
stimuli at all (Fig.\ref{fig:reciprocity}). The network performance
is, however, quite robust: the overall decorrelation deteriorates
noticeably  only when more than 50\% of the connections have been
rewired. The highly correlated stimuli are, however, more sensitive
to the rewiring with $r^{(top)}$ increasing from $r^{(top)}=0.44$
to $r^{(top)}=0.52$ when 50\% of the connections are rewired, while
$\bar{r}$ changes only from $\bar{r}=-0.08$ to $\bar{r}=-0.05$.

\begin{figure}[!h]
\begin{center}
\includegraphics[width=14cm]{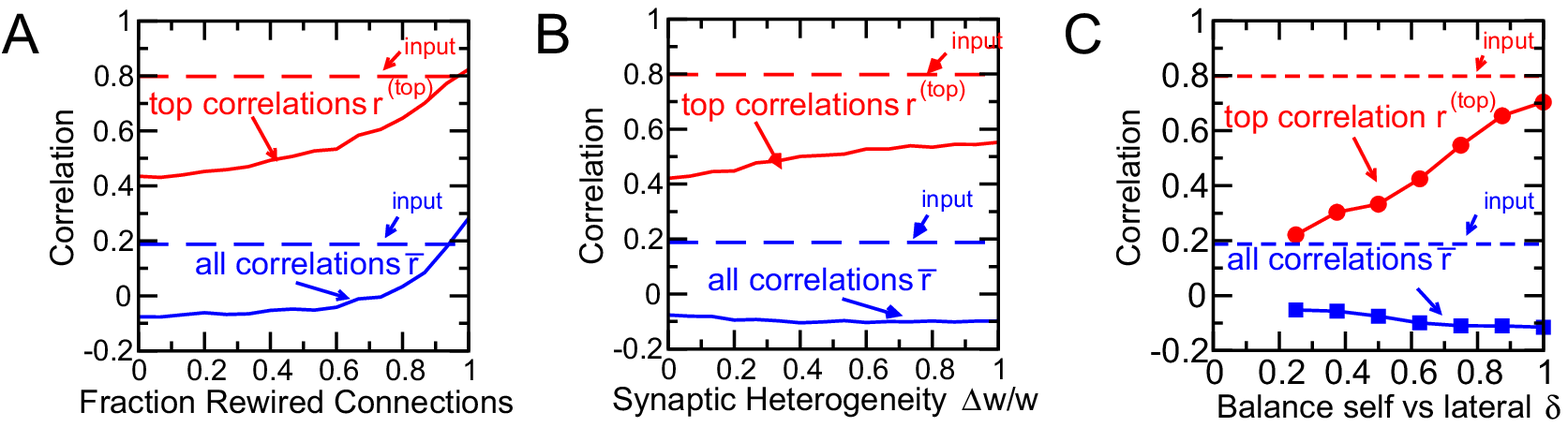}
\end{center}
\caption{
{\bf Effective decorrelation does not require complete reciprocity of the
synapses.}\textbf{ A)} A fraction of the inhibitory connections are
rewired to a randomly chosen mitral cell. Dashed lines denote input
correlations.\textbf{ B)} The inhibitory synaptic strengths are picked
with equal probability from the two values $w\pm\Delta w$. \textbf{C)
}Reducing self-inhibition in favor of lateral inhibition, $\delta<\frac{1}{2}$,
enhances the decorrelation. Parameters: $\gamma=10$, $R_{0}=0.1$,
$G_{min}=1.2$, $N_{m}=424$, $n_{connect}=8$, $\beta=33$, $w=0.005$,
$M_{sp}=1$.\textbf{ }
}
\label{fig:reciprocity}
\end{figure}

The granule cells deliver their inhibitory inputs onto the secondary
dendrites of the mitral cells at highly variable distances from the
mitral cell somata. Their effect on the mitral cell firing will therefore
vary over quite some range; in fact, some synaptic contacts will be
too far away from the mitral cell soma to have any noticeably effect
on that mitral cell's firing. To assess the impact of such heterogeneities
we modify the inhibitory synaptic weights, which so far had the same
value $w$ for all synapses, by picking them with equal probability
from the two values $w\pm\Delta w$. This breaks the symmetry of the
inhibition and for $\Delta w=w$ half of the inhibitory connections
are completely ineffective. The overall decorrelation is, however,
not affected by this heterogeneity and even the decorrelation of the
highly similar stimuli deteriorates only slightly over the whole possible
range $0\le\Delta w\le w$ (Fig.\ref{fig:reciprocity}B). Essentially
the same result is obtained if the synaptic strengths are distributed
uniformly in the interval $[w-\Delta w,w+\Delta w]$. While for very
large granule cell populations the heterogeneities of different granule
cells are expected to average out each other, for the parameters used
in our study the effective connectivity matrix is still noticeably
asymmetric: its anti-symmetric component amounts to about 20\% of
the symmetric one.

Through the reciprocal character of the dendrodendritic synapse a
granule cell mediates lateral inhibition between the mitral cells
that drive it as well as self-inhibition of each of them. Due to the
complex dendritic dynamics of granule cells \cite{EgUr06,Eg08} these
two types of inhibition can be of different strength. In fact, recent
observations suggest that self-inhibition is significantly weaker
than lateral inhibition \cite{DavSch11}. While our minimal model
does not capture any explicit dendritic processing, the strength of
self-inhibition and lateral inhibition that a mitral cell receives
is given by the diagonal and off-diagonal coefficients of the effective
connectivity matrix $\boldsymbol{\mathcal{W}}\equiv\mathbf{W}^{(mg)}\mathbf{W}^{(gm)}$,
respectively. We can therefore change the balance between self-inhibition
and lateral inhibition phenomenologically by rescaling the diagonal
terms, $\mathcal{W}_{ii}\rightarrow\mathcal{N}^{-1}\delta\mathcal{W}_{ii}$
with $0\le\delta\le1$, at the expense of the off-diagonal terms,
$\mathcal{W}_{ij}\rightarrow\mathcal{N}^{-1}(1-\delta)\mathcal{W}_{ij}$
for $i\ne j$, while keeping the row-sum of the matrix fixed through
the normalizing factor $\mathcal{N}$.\textbf{ }Reducing self-inhibition
in this fashion ($\delta<\frac{1}{2}$) enhances the decorrelation
of the representations of the natural stimuli significantly (Fig.\ref{fig:reciprocity}C),
because it further enhances the competition between dominant, co-active
mitral cells. Conversely, increasing the self-inhibition ($\delta>\frac{1}{2}$)
reduces the competition. In the complete absence of lateral inhibition
($\delta=1$) granule cells are effectively coupled only to a single
mitral cell. This provides still good overall decorrelation, but the
representations of the highly similar odors are only poorly decorrelated.
Thus, the experimentally observed reduction of self-inhibition may
contribute to an improved decorrelation performance of the bulbar
network.

These comparisons show that for effective decorrelation the most important
aspect of the reciprocity of the dendrodendritic synapse is that it
provides mutual anatomical connections between the relevant mitral
cells, i.e. between those that are co-active for some stimuli. The
effective synaptic strengths can be quite heterogeneous without compromising
the performance of the network. In fact, reduced self-inhibition can
enhance the decorrelation substantially.

\subsection*{Young Granule Cells Respond to Novel Odors}

One possible role of neurogenesis is to provide a persistent supply
of new neurons, which may play a different role than old, mature neurons.
An aspect of this type has been identified in experiments focusing
on the responsiveness of young and old adult-born granule cells \cite{MaMi05,BelKoe11}.
In the experiments, adult-born precursor cells, which develop into
granule cells, were marked in the subventricular zone. After they
have migrated to the olfactory bulb and have integrated into the bulbar
network their response to odor stimulation was measured using the
expression levels of various immediate early genes. It was found that
the fraction of adult-born granule cells that respond to novel odors
is significantly higher shortly after their arrival in the olfactory
bulb\textbf{ }than a few weeks later. It has been argued therefore
that one important function of the young granule cells may be to serve
as novelty detectors \cite{MaMi05}.

In our computational model a differential response of young and older
adult-born granule cells to novel odors arises quite naturally. After
establishing a network by exposing the system to the stimulus ensemble
$\mathbf{S}^{(\alpha)}$, $\alpha=1,\ldots,7$, we mark granule cells
as they are integrated into the network and measure their response
to various stimuli as a function of their age. Assuming that the granule
cell activity has to surpass a minimal value to activate the expression
of the immediate early genes, we consider granule cells as responding
if they reach an activity above a threshold $G_{IEG}$. As the network
evolves the less active granule cells die and are removed from the
network (Fig.\ref{fig:novel-odors}Ai). As in the experiments, we
find that the fraction of young adult-born granule cells that respond
to a novel stimulus, i.e. a stimulus that is quite different from
the stimuli in the background ensemble, decreases as the granule cells
become older (Fig.\ref{fig:novel-odors}Aii). This decrease results
from the reduced survival probability of these cells, which is due
to the weak drive they receive by the stimuli in the stimulus ensemble
that determines granule-cell survival. In contrast, the fraction of
granule cells that respond to a familiar stimulus, i.e. a stimulus
in the background ensemble, decays very little or even increases over
the same time frame, reflecting their higher survival rate. 

\begin{figure}[!h]
\begin{center}
\includegraphics[width=14cm]{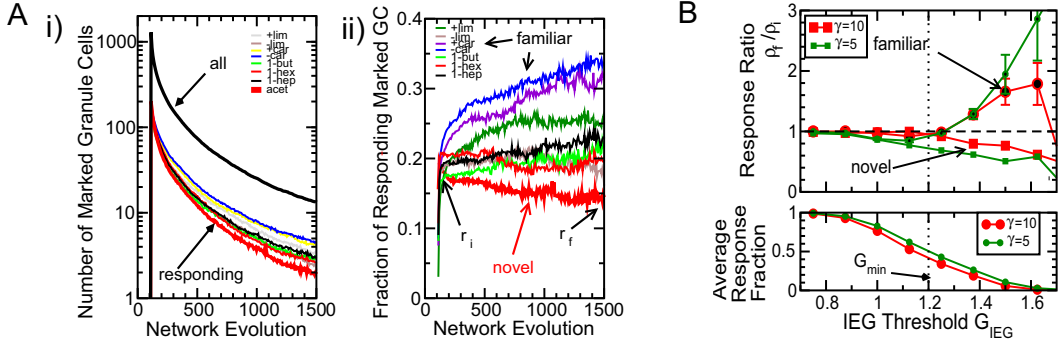}
\end{center}
\caption{
{\bf Young granule cells show enhanced response to novel odors.} \textbf{A)
i) }Granule cells are marked at $t=100$. The total number of marked
cells (black thick line) and the number of marked cells responding
to one of the eight stimuli decreases with time. The stimulus ensemble
consists of $\mathbf{S}^{(\alpha)}$, $\alpha=1\ldots7$. Stimulus
8 (acetic acid) is novel (cf. Fig.\ref{fig:mouse_stimuli_decorrelation}).
\textbf{ii)} The fraction $\rho(t)$ of marked granule cells that
respond to the novel stimulus decreases with time. For the familiar
stimuli it mostly increases. Parameters: $\gamma=10$, $R_{0}=0.1$,
$G_{min}=1.2$, $N_{m}=424$, $n_{connect}=8$, $\beta=33$, $w=0.005$,
$M_{sp}=1$, $G_{IEG}=1.375$.\textbf{ B) }The IEG-activation threshold
$G_{IEG}$ has to be close to the resilience threshold $G_{min}$.
Bottom panel: for $G_{IEG}$ well above $G_{min}=1.2$ (dotted line)
very few marked cells reach an activity above $G_{IEG}$ and are considered
as responding to the stimuli. Top panel: For $G_{IEG}>G_{min}$ the
response fraction $\rho$ decreases with time for the novel stimulus
(ratio of response fractions $\rho_{f}/\rho_{i}<1$ , cf. panel Aii),
while it tends to increase for the familiar stimuli ($\rho_{f}/\rho_{i}>1$,
error bars denote standard deviation across the stimuli $\mathbf{S}^{(\alpha)}$,
$\alpha=1,\ldots,7$).\textbf{ }Parameters as in \textbf{A }except
for the steepness $\gamma$ of the survival curve (cf. Fig.\ref{fig:network_sketch}Bi),
$\gamma=5$ (green, small symbols), $\gamma=10$ (red, large symbols).
The results represent an\textbf{ }average across 32 runs.\textbf{ }
}
\label{fig:novel-odors}
\end{figure}

For what range of the threshold $G_{IEG}$ does our model yield results
that agree qualitatively with the experiments in \cite{MaMi05}? When
the threshold $G_{IEG}$ is increased beyond the resilience threshold
$G_{min}$ ever fewer marked granule cells respond and the fraction
of marked granule cells that respond to the stimuli - averaged over
all stimuli - drops from 1 to 0 (Fig.\ref{fig:novel-odors}B bottom
panel). Thus, the experimentally obtained response fractions of 10-20\%
\cite{MaMi05} set an upper limit for $G_{IEG}$ relative to $G_{min}$.
At the same time, decreasing $G_{IEG}$ reduces the difference between
the temporal evolution of the response to novel and to familiar stimuli.
We characterize the evolution by the ratio $\rho_{f}/\rho_{i}$ between
the fraction $\rho_{f}$ of granule cells responding to the stimulus
at the final time of the simulation and the fraction $\rho_{i}$ immediately
after the end of the marking period. On average this ratio increases
with increasing $G_{IEG}$ for the familiar stimuli, but it decreases
for the novel stimulus (Fig.\ref{fig:novel-odors}B top panel). For
the response to novel odors to differ significantly from that to familiar
odors $G_{IEG}$ cannot be much smaller than $G_{min}$. It is worth
noting that varying the steepness $\gamma$ of the survival curve
does not affect the decorrelation of the odor stimuli substantially\textbf{
}(symbols at $t=1,450$ in the\textbf{ }top panel of Fig.\ref{fig:network_evolution}),
but the difference in the response to novel compared to familiar odors
is significant only if the survival curve is not too steep (Fig.\ref{fig:novel-odors}B
top panel). 

Thus, the activity-dependent survival of the granule cells combined
with their random connections to the mitral cells is sufficient to
capture the experimentally observed enhanced response of young adult-born
cells to novel stimuli if the threshold $G_{IEG}$ for the activation
of the immediate early genes is close to the resilience threshold
$G_{min}$, which is an essential determinant of the survival of the
granule cells. 

\subsection*{Neurogenesis Contributes to Perceptual Learning}

In a wide range of experiments possible connections between adult
neurogenesis and animal performance have been investigated employing
various tests of odor detection, odor discrimination, short-term and
long-term memory, and fear conditioning \cite{RoGh02,ImSa08,BrLe09,MoLe09,SuMa10,LaMo09,VaMu09,MaSt06,MaSt06a,MaSt06b,MoLi09}.
No simple picture regarding the role of neurogenesis in odor discrimination
and odor memory has, however, emerged so far. This may in part be
due to the fact that higher brain areas are likely involved in many
of the behavioral tasks; they may well compensate for some changes
occurring in the olfactory bulb and therefore possibly mask certain
effects of the neurogenesis.

A behavioral task that may reflect bulbar odor representations
relatively directly is the spontaneous, non-associative odor
discrimination based on habituation, which has been shown to result
predominantly from bulbar processes \cite{McMa08,WiLi08,ChMa10}. These
experiments exploit the decreasing interest an animal typically
displays to repetitions of the same stimulus: the animal's response to
a second stimulus after if has habituated to a first stimulus is a
measure of the degree to which the animal discriminates the two
stimuli \cite{WiLi08}. Exposing animals to extended periods during
which their environment is enriched with additional odors enhances
their spontaneous odor discrimination
\cite{MaSt06,MaSt06a,MaSt06b,MoLi09}. This is indicative of perceptual
learning. The dominance of bulbar processing in this task
\cite{McMa08,WiLi08,ChMa10} suggests that the enrichment induces
changes in the bulbar odor representations \cite{ChMa10}. Since the
enhancement is significantly suppressed if neurogenesis is halted
pharmacologically \cite{MoLi09}, it is likely that the changes in the
odor representations reflect a restructuring of the bulbar network.
Importantly, for the enrichment to improve the performance the odors
employed have to be related to the odors that are to be discriminated
\cite{MaSt06}.

\begin{figure}[!h]
\begin{center}
\includegraphics[width=14cm]{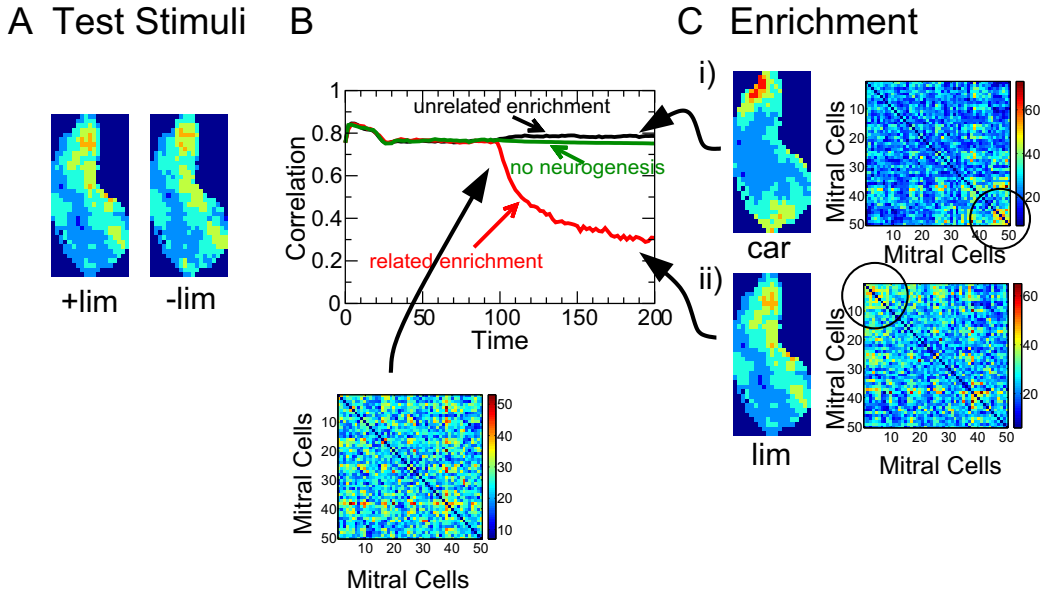}
\end{center}
\caption{
{\bf Perceptual learning.} Correlation \textbf{(B)} of the test stimuli\textbf{
}$+$-limonene and $-$-limonene\textbf{ (}shown in \textbf{A) }as
a function of time. Enrichment, beginning at $t=100$, changes the
connectivity. Enrichment with the related odors\textbf{ }$+$-limonene
and $-$-limonene\textbf{ }(\textbf{Cii}, only $+$-limonene is shown)
strongly reduces the correlation, whereas enrichment with the unrelated
odors $+$-carvone and $-$-carvone (\textbf{Ci}, only $+$-carvone
shown) does not. Enrichment with a related odor but without neurogenesis
does not enhance the decorrelation. Parameters: $\gamma=20$, $R_{0}=0.1$,
$G_{min}=1.2$, $N_{m}=424$, $n_{connect}=8$, $\beta=33$, $w=0.005$,
$M_{sp}=1$.\textbf{ }Background stimuli: 1-butanol, 1-hexanol, 1-heptanol,
and acetic acid. 
}
\label{fig:enrichment}
\end{figure}

The perceptual learning observed in the experiments is captured in
our minimal computational model. We use an ensemble of background
stimuli,\textbf{ }which establishes a default network connectivity,
and test the performance of the network with two test stimuli ($+$-limonene
and $-$-limonene). They are not included in the stimulus ensemble
that drives the network evolution. For the default network the correlation
between the representations of the test stimuli is high, consistent
with the fact that naive animals do not discriminate these odors spontaneously.
Then the stimulus ensemble is enriched with additional odors ($t>100$).
As the network adapts and evolves to a new steady state characterized
by different effective connectivity matrices (Fig.\ref{fig:enrichment}C,
right panels), the correlation between the two test stimuli evolves
as well. If the odors used for the enrichment have sufficient overlap
with the test odors the correlation between the test odors decreases
substantially (red line in Fig.\ref{fig:enrichment}B). However, if
the enrichment odors are only weakly related to the test odors the
correlation of the test odors does not decrease (black line in Fig.\ref{fig:enrichment}B).
In fact, in some cases the correlation between the test odors can
even increase. As expected, if the influx of new granule cells is
stopped with the onset of the enrichment the odor representations
and their correlations are unaffected by the enrichment, even if the
enrichment odor is close to the test odor (green line).

\subsection*{Effective Enrichment: Overall Overlap is Not Sufficient}

In experiments, odor enrichment enhances the ability of the animals
to discriminate similar odors only if there is sufficient overlap
between the activation patterns of the stimuli used in the enrichment
and those of the stimuli to be discriminated \cite{MaSt06}. Our network
model allows more specific predictions for the type of enrichment
protocols that should be effective in enhancing the ability of the
animals to discriminate a given set of test odors.

We consider the decorrelation of very similar mixtures comprised of
dissimilar components. Specifically, we use as components limonene
(50\% $+$--limonene and 50\% $-$--limonene) and carvone (also both
enantiomers in equal proportions), whose activation patterns have
very little overlap (Fig.\ref{fig:mouse_stimuli_decorrelation}A).
We employ two different enrichment protocols. In the first one pure
limonene and pure carvone are added to a background of alcohols and
acetic acid in an alternating fashion (Fig.\ref{fig:protocol}B, top
panel). Experimentally, this would correspond to presenting limonene
and carvone separately at different times. In the second protocol
an equal mixture of limonene and carvone is added to the background
ensemble (Fig.\ref{fig:protocol}B, bottom panel). In both protocols
the activity-dependent removal of interneurons occurs only after the
complete set of background and enrichment stimuli has been presented.
To implement the mixtures in the model we assume that the glomerular
activation patterns for mixtures are approximated sufficiently well
by a linear combination of the patterns for the individual components. 

\begin{figure}[!h]
\begin{center}
\includegraphics[width=14cm]{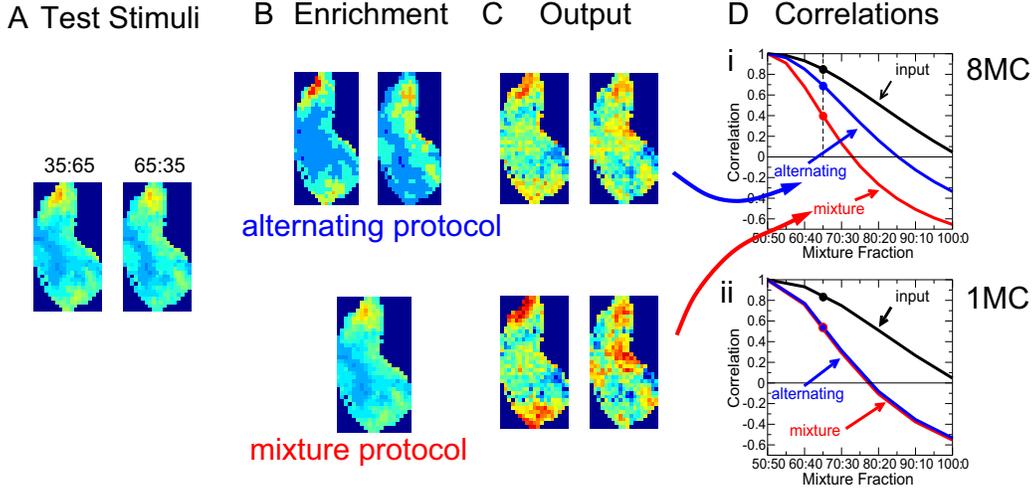}
\end{center}
\caption{
{\bf Effect of enrichment protocol on the decorrelation of similar mixtures.}
\textbf{A)} Sample test stimuli: mixtures of $\pm$--limonene and
$\pm$--carvone with mixture fractions 35:65 and 65:35. \textbf{B)}
Enrichment stimuli. i) $+$-limonene, $-$-limonene, $+$-carvone,
and $-$-carvone alternating (only $+$-carvone and $+$-limonene
shown), ii) 50:50 mixture of $\pm$-limonene and $\pm$-carvone. \textbf{C)}
Output patterns for the test stimuli shown in A. Parameters as in
Di. \textbf{D)} Correlations of mitral cell activities for the test
stimuli as a function of the mixture fraction. i) Eight connections
per granule cell. The mixture protocol achieves substantially better
decorrelation than the alternating protocol. Parameters: $\gamma=20$,
$R_{0}=0.1$, $G_{min}=1.2$, $N_{m}=424$, $n_{connect}=8$, $\beta=33$,
$w=0.005$, $M_{sp}=1$.\textbf{ }ii) One connection per granule cell.
No significant difference between the protocols.  Parameters: $\gamma=20$,
$R_{0}=0.1$, $G_{min}=0.1275$, $N_{m}=424$, $\beta=0.125$, $w=0.3$,
$M_{sp}=1$. 
}
\label{fig:protocol}
\end{figure}

While using the pure components in the enrichment decreases the correlation
between the representations of the mixtures at all mixture ratios
(`alternating' in Fig.\ref{fig:protocol}Di) it does so substantially
less than the network enriched with the 50:50 mixture (`mixture' in
Fig.\ref{fig:protocol}Di). The stronger decorrelation obtained with
the mixture protocol compared to the alternating protocol can also
be recognized directly in the output patterns (Fig.\ref{fig:protocol}C,
bottom vs. top panel). This substantial difference arises because
the decorrelation of the mixtures is strongly enhanced by mutual inhibition
between mitral cells that are driven by limonene as well as carvone.
This inhibition is provided by `mixed' granule cells.  By contrast,
`pure' granule cells are connected to mitral cells that are activated
(almost) exclusively by limonene \emph{or} carvone. As discussed in
Sec.~\textit{Threshold Promotes  Lateral Inhibition Based on Co-Activity}, in the context of interference, in
the alternating protocol the mixed granule cells have a lower  survival
probability  than the pure granule cells. In the mixture protocol,
however, both types of granule cells have very similar survival probabilities.
To assess the inhibition provided by these populations we consider
the sum of the synaptic weights in the four quadrants of the effective
connectivity matrix $\boldsymbol{\mathcal{W}}$ (cf. Fig.\ref{fig:network_evolution}Biii).
We find that the inhibition provided by the mixed granule cells in
the mixture protocol is stronger than in the alternating protocol
(see (\ref{eq:blocks_of_populations})). Insight into what controls
these populations can be gained by considering again the simple caricature
of Fig.\ref{fig:Phase-plane-analysis}A. Within that framework the
mixture protocol can be viewed as a stimulus set in which all four
glomeruli receive essentially equal input. Both types of granule cells
have then equal survival rates. Within that model it is easily seen
that the size of their populations falls between that of the mixed
granule cells and the pure granule cells in the alternating protocol
because the total resilience of the mixed granule cells has to be
the same in both protocols  (see Sec.~\textit{METHODS, Alternating vs Mixture Protocol}).
Thus, compared to the alternating protocol the mixture procotol enhances
the relevant inhibition and improves the decorrelation of the limonene-carvone
mixtures.

If neurogenesis were to affect only interneurons that provide non-topographic
inhibition and no lateral inhibition \cite{ClSe06} both enrichment
protocols would be expected to lead to the same level of decorrelation.
Specifically, if in the model each granule cell makes only connections
with a single mitral cell the alternating protocol leads to the same
decorrelation of the limonene-carvone mixtures as the mixture protocol
(Fig.\ref{fig:protocol}Dii). Comparing the influence of the two enrichment
protocols on the animals' ability to discriminate such mixtures may
therefore give insight into the type of neurogenesis-dependent connectivity
that dominates the decorrelation mechanism.

Thus, even though in both protocols the enrichment odors - taken together
- have the same overlap with the test odors the model predicts that
enrichment with the mixture protocol achieves substantially better
decorrelation of the test stimuli than the alternating protocol.

\section*{Discussion}

To investigate the functional implications of the experimentally observed
persistent turnover of inhibitory interneurons on sensory processing
by the olfactory bulb we have used a minimal computational network
model. The experimental observations forming the basis of our model
are the reciprocity of the synapses between the interneurons (granule
cells) and the principal neurons (mitral cells) \cite{Sh04} and the
activity-dependent survival of adult-born granule cells \cite{LiSi10}.
In the model we have focused on the input from the mitral cells via
the dendro-dendritic synapses as the dominant input controlling the
activity and survival of the granule cells. Assuming in addition that
the new cells connect to an essentially random set of mitral cells
allows the model to capture parsimoniously various experimental observations
and to make specific predictions. 

\textbf{Novelty Detection. }It has been observed that young granule
cells are more likely than mature ones to respond to odors that are
novel for the animal \cite{MaMi05,BelKoe11}. This has been interpreted
as a mechanism for novelty detection. Our model captures the enhanced
response of young cells in a natural way. Since granule cells that
respond to novel odors but not to the odors in the ongoing environment
receive only little ongoing input, they do not survive for a long
time and the fraction of granule cells responding to the novel odor
decreases with their age. Thus, the observation of an enhanced response
of young granule cells to novel odors suggests that new granule cells
do not have a strong bias towards connecting to highly active mitral
cells but connect also to mitral cells that have only been weakly
active in the past. Such a strategy enables the network to learn to
process novel odors. 

Experimentally, the response of the granule cells was measured in
terms of the expression of various immediate early genes (c-fos, c-jun,
EGR-1/zif-268). The fraction of granule cells responding to the novel
odors was found to be 10-25\% for young cells and lower for older
cells \cite{MaMi05,BelKoe11}.\textbf{ }Such an intermediate response
fraction is obtained in our model if the threshold for the expression
of the immediate early genes is close to that for the survival of
the granule cells. This is suggestive of a common step in the pathways
controlling IEG-expression and cell survival. %

\textbf{Threshold Enhances Inhibition between Co-active Mitral Cells
and Reduces Interference. }The decorrelation of highly similar stimuli
like the two pairs of enantiomers used in our computation hinges upon
the presence of an activity threshold that the granule cells have
to surpass to increase their survival probability. It  enhances the
connections between mitral cells that are highly active simultaneously
and suppresses those between mitral cells that are strongly active
albeit only in response to different stimuli.

Biophysically, a threshold for the survival of the granule cells may
arise from the need to drive L-type Ca channels, which activate the
MAPK pathway that leads to the stimulation of genes that are essential
for neuronal survival \cite{DoPa01,MiSt05}. 

With the strengthening of inhibition between co-active mitral cells
the mechanism underlying the adaptation in our model is somewhat related
to that underlying other adaptive networks that have been studied
previously. In an early neurogenesis model for the olfactory bulb
the evolution of the effective pairwise inhibition between mitral
cells was based directly on the scalar product of the mitral cell
activities \cite{CePe01}. Adaptive networks that aim to orthogonalize
the stimulus representations can do so via a connectivity that is
based on the pairwise scalar products of input activities \cite{WiWi10}.\textbf{
}A somewhat different adaptive connectivity has been suggested in
a modeling study of the bee antennal lobe. There it was found that
a connectivity in which the inhibition is proportional to the correlations
between the glomerular activities was able to match the observed output
patterns better than random or local center-surround connectivities
\cite{LiSa05}. We have compared a few types of networks that exploit
different adaptation algorithms and find that connectivities that
are based on the co-activity of mitral cells or glomeruli achieve
significantly better decorrelation than networks based on the correlations
or covariances of the inputs. A particular problem of the latter algorithms
is that they are not sensitive to mean activities of the cells and
do not take the spontaneous activity of the mitral cells adequately
into account.

\textbf{Reciprocity of Connections. }An anatomically characteristic
feature of the olfactory bulb is the reciprocal nature of the dendrodendritic
synapses between mitral and granule cells. The purpose of this reciprocity
is not well understood. Our computational modeling shows that it can
play an essential role in exploiting the activity-dependent survival
of the granule cells to establish a connectivity whose lateral inhibition
reflects the co-activity of the mitral cells.  This provides a mechanism
for the network to learn to decorrelate even highly similar stimuli. 

{Biologically, the reciprocity may be imperfect in a number of ways.
In principle, an inhibitory synapse could connect the granule cell
to a mitral cell that is not the origin of the associated excitatory
synapse. Modeling such a situation by a random rewiring of a fraction
of inhibitory connections we find that the network performance is
reasonably robust to such perturbations. However, when more than 50\%
of the synapses are rewired the performance deteriorates significantly
and without any reciprocity the stimulus representations are not decorrelated
at all.

A second type of imperfection of the reciprocity is likely to arise
if the dendrodendritic synapse is located far from the soma of the
mitral cell. In such a case the inhibition exerted by the granule
cell may not have much effect on the mitral cell firing, although
the granule cell is driven strongly by that mitral cell. This asymmetry
can arise because excitation is driven by action potentials, which
can travel long distances along the dendrite, whereas the shunting
provided by the inhibition is confined to a distance comparable to
the electrotonic length of the dendrite \cite{DaLi08}. Thus, the
effective inhibitory strength may vary substantially between synapses
depending on their location relative to the soma. Mimicking such a
heterogeneity by random variations in the synaptic strength we find
that the network performance is only moderately affected by such effects.
Since mitral cells are connected to many granule cells the heterogeneity
of the combined synaptic strengths is likely to be reduced compared
to the heterogeneities within individual granule cells. Such an averaging
may be reduced if correlations between the strengths of different
synapses, which may arise due to correlations in the physical distances
between the cells, should be significant. 

The reciprocity may also be perturbed because the strength of the
self-inhibition that a mitral cell experiences on account of a given
granule cell may differ from that of the lateral inhibition that said
granule cell provides to other mitral cells. In fact, recent experiments
suggest that self-inhibition is significantly weaker than lateral
inhibition \cite{DavSch11}. One cause for this difference may be
the complex physiology of granule cells, which includes local dendritic
calcium signaling, dendritic calcium spikes, and action potentials
driven by sodium conductances \cite{Eg08}. Our minimal single-compartment
model for the granule cell does not allow to capture these rich dynamics.
However, on a phenomenological level the balance between self-inhibition
and lateral inhibition can be modified by rescaling the diagonal and
off-diagonal terms in the effective connectivity matrix. Our model
shows that reducing the self-inhibition while strengthening the lateral
inhibition can substantially enhance the ability of the network to
decorrelate the representations of highly similar stimuli. 

\textbf{Perceptual Learning. }The decorrelation of similar stimulus
representations that is obtained in our model provides a natural interpretation
of recent experiments on spontaneous odor discrimination via habituation
\cite{MoLi09}. Only with neurogenesis intact does enriching an animal's
odor environment enhance its ability to discriminate similar odors.
Since the habituation used in these discrimination experiments reflects
predominantly changes in the olfactory bulb rather than higher brain
areas \cite{McMa08,WiLi08,ChMa10}, the improvement in odor discrimination
resulting from odor enrichment likely reflects modifications in the
encoding of the test stimuli in the olfactory bulb. Our modeling shows
that fundamental features underlying the neuronal turn-over in the
bulb -- activity-dependent survival and reciprocal synapses -- suffice
to allow perceptual learning by changing the odor encoding so as to
decrease their similarity and enhance their discriminability. %

\textbf{Repertoire of Potentially Relevant Odors. }In laboratory experiments
that allow many repetitions animals can learn to discriminate highly
similar odor stimuli \cite{MoLe09}, which may have highly correlated
representations in the olfactory bulb. Outside the laboratory the
animals are likely to face the challenge to form associations with
stimuli given only a few trials. This task may be very difficult if
not even impossible if the relevant odors are represented in the bulb
in a highly correlated fashion. 

In line with experiments on odor enrichment \cite{MoLi09}, our computational
model shows that neurogenesis may facilitate this task by reducing
the correlation of odors in an ensemble to which the animal is exposed.
These odors could represent a repertoire of potentially relevant odor
types that the animal can easily discriminate, should the need arise.
In our model the survival of the granule cells depends on the inputs
they receive from mitral cells via their dendro-dendritic synapses.
Their relevance could be determined by the context in which the animal
is exposed to the odor. Such contexts are likely to affect modulatory
inputs to the olfactory bulb, which can modify the excitability of
granule cells \cite{CaCa99,PrIn07,EsAr10} and mitral cells \cite{CaCa99,EsAr10}
as well as mitral cell inhibition \cite{NaDo09}, all of which will
affect granule cell survival \cite{KaOk06}. Contexts could also induce
specific, direct inputs from cortical areas like piriform cortex to
granule cells at proximal or basal synapses, both of which are functional
in young granule cells with the proximal synapses developing even
before the basal and dendro-dendritic ones \cite{WhGr07,KeLi08}.

\textbf{Predictions for Effective Learning Protocols. }Enrichment
enhances odor discrimination only if the enrichment odors overlap
in their glomerular excitation patterns with those of the test stimuli
\cite{MaSt06}. Our modeling confirms this. Moreover, it makes specific
predictions with regard to the decorrelation of similar mixtures comprised
of dissimilar components. If neurogenesis affects predominantly granule
cells that provide lateral inhibition, our model predicts that animals
will learn to discriminate such mixtures more easily if the enrichment
is performed using the odor mixture rather than alternating its individual
components. This difference is predicted even though both enrichment
ensembles have the same overall overlap with the test stimuli. If
non-topographical self-inhibition \cite{ClSe06} should dominate neurogenetic
restructuring, no difference between the protocols is expected. 

The change of odor representations that our neurogenesis model predicts
to arise from odor enrichment might also be testable in reward-associated
discrimination tasks by focusing on the initial learning stages. Suitable
enrichment protocols are expected to enhance the differences in the
encoding of the test odors. Applied before the animals learn the odors
that are to be discriminated, such enrichment should lead to a shortening
of the initial learning phase if the odors are very similar. Moreover,
it has been found that animals tend to follow different strategies
during the early stages of a 2-alternative choice odor discrimination
task depending on the degree of similarity of the two odors \cite{KaBe10}.
In fact, for very similar stimuli their early strategy suggests that
they actually can not yet tell the test odors apart. In that case
suitable prior enrichment may even allow the animals to employ their
coarse-discrimination strategy for odor pairs for which without enrichment
they would use the fine-discrimination strategy.%

\textbf{Decorrelation by Individual and Joint Normalization.} Divisive response normalization has been discussed extensively in
sensory processing, in particular in the visual system \cite{He92}.
In this type of normalization the response of each cell, which corresponds
to a channel with given response characteristics like preferred orientation
or spatial frequency of visual grid patterns, is divided by the sum
of the activities of cells covering a wider range of response characteristics.
The gain control implemented by this process is consistent with various
experimentally observed neural responses (e.g. contrast independence
and contrast adaption) \cite{He92}. In olfaction it has been proposed
that such a normalization may arise from the lateral inhibition provided
by the network of peri-glomerular cells, short-axon cells, and external
tufted cells in the glomerular layer of the olfactory bulb \cite{ClJo07}.
Divisive normalization has been observed in the antennal lobe, which
is the insect analogue of the olfactory bulb \cite{OlBh10}. Implemented
in simulations by global lateral inhibition, it was found to reduce
the correlation between the different channels (activities of the
principal neurons) across a large set of odors \cite{OlBh10,LuAx10}. 

Further analysis of our neurogenetic model suggests that the olfactory
bulb performs a complementary type of divisive normalization (unpublished data).
Rather than reshaping the mitral cell activities such that their pattern
average is the essentially the same for all stimuli, the activity-dependent
survival of the granule cells tends to equalize (normalize) the activity
of all mitral cells when averaged across the stimulus set. Correspondingly,
it foremost contributes to a reduction of the correlations between
pairs of activity patterns rather than between pairs of mitral cells
(channels). For stimuli whose similarity is dominated by highly co-active
mitral cells the normalization of the activity of individual mitral
cells achieves, however, only quite limited decorrelation. The joint
normalization of the activities of multiple mitral cells, which results
from the lateral inhibition of granule cells connected to multiple
mitral cells, can preserve some differences in the mitral cell activities
and, as a consequence, can achieve considerably better pattern decorrelation. 

\textbf{Limitations of the Model.} In our minimal model we have focused
on the impact of the structural plasticity afforded by the turn-over
of the granule cells. We have therefore treated the individual mitral
and granule cells in a minimalistic fashion. In particular, we have
described them in terms of linear  rate dynamics  without any threshold.
Previous studies have shown that nonlinearities can induce stimulus
decorrelation even in non-adaptive networks \cite{ClSe06,ArKa08,WiJu10}.
An interesting question is therefore whether neuronal nonlinearities
could further enhance the decorrelation achieved by the adaptive networks
studied here. 

Moreover, we have modeled each neuron as a single compartment. Both,
mitral and granule cells, have, however, elaborate dendrites, which
most likely increase the complexity of their interaction. Thus, while
action potentials can propagate with little attenuation along the
mitral cell dendrite and can excite granule cells even at large distances,
the inhibition that a granule cell imparts to a mitral cell is expected
to depend strongly on the distance of the GABAergic synapse from the
mitral cell soma. The mutual inhibition that a granule cell mediates
between mitral cells will then not be symmetric. We have mimicked
such an asymmetry by modifying the effective connectivity matrix and
found that the network performance is quite robust with respect to
such perturbations. 

The dendritic computations in the granule cells are tied in with their
complex multi-level calcium dynamics \cite{EgUr06,Eg08}. Even quite
small depolarizations of a granule cell spine can induce local GABA
release, which results in graded self-inhibition of the driving mitral
cell. Stronger inputs can induce low-threshold calcium spikes that
can spread within the dendritic tree. Finally, suitable inputs can
trigger somatically evoked conventional sodium-driven action potentials
that invade the whole dendrite. This complexity may endow the granule
cell with additional computational power like a dynamically regulated
range of inhibition. Since the ability to generate sodium spikes develops
last in adult-born granule cells \cite{LlAl06}, the balance between
local signaling, calcium spikes, and sodium spikes may change with
the age of the cell. While our phenomenological modeling does not
capture this biophysical complexity, it shows that a reduction of
the self-inhibition and a concomitant enhancement of lateral inhibition
can substantially improve the decorrelation of stimulus representations. 

The mechanisms controling the survival and apoptosis of granule cells
are not understood in detail. It is known that larger fractions of
granule cells survive if the animal is kept in odor-enriched enviroments
\cite{RoGh02} or if the excitability of the granule cells is genetically
enhanced \cite{LiSi10}. In our minimal model we therefore assumed
that the survival of the granule cells increases with their activity.
It has been found, however, that certain associative odor discrimination
tasks can not only enhance but also reduce the survival of the adult-borne
granule cells, depending on their age \cite{MoGh08}. Recent experiments
have also indicated that apoptosis of specific neurons can be elevated
when associative memories are erased \cite{SulDid11a}. It would be
interesting to extend our minimal model, which aims to capture the
impact of neurogenesis on non-associative odor discrimination tasks
\cite{MaSt06a,MaSt06b,MoLi09}, to such more complex situations. 

In conclusion, using a minimal computational model we have shown that
adult neurogenesis with activity-dependent apoptosis of inhibitory
interneurons that are reciprocally connected with the principal neurons
is sufficient to restructure a network like that of the olfactory
bulb such that it learns to decorrelate representations of very similar
stimuli. The network performance is quite robust with respect to various types
of deviations from reciprocity that are likely to be present in the
olfactory bulb. The model makes predictions regarding the impact of
different enrichment protocols on the performance of animals in spontaneous
and award-related odor discrimination tasks. Their outcome is expected
to give insight into the type of network connectivity that is associated
with the interneuronal turn-over. 

% You may title this section "Methods" or "Models". 
% "Models" is not a valid title for PLoS ONE authors. However, PLoS ONE
% authors may use "Analysis" 
\section*{Methods}

\subsection*{Discrete Adaptive Network Model}

We consider a minimal computational network model that focuses on
the turn-over of inhibitory interneurons caused by neurogenesis and
activity-dependent apoptosis and study the networks' ability to learn
to decorrelate similar stimulus representations. The recurrent network
comprises two types of neurons, principal neurons (mitral cells) and
inhibitory interneurons (granule cells), which are coupled through
reciprocal synapses (Fig.\ref{fig:network_sketch}). Within the framework
of threshold-linear rate equations the activity $M_{i}^{(\alpha)}$
of mitral cell $i$ in response to stimulus $\alpha$ with afferent
activity pattern $S_{i}^{(\alpha)}$, $i=1\ldots N_{m}$, and the
corresponding activity $G_{j}^{(\alpha)}$ of granule cell $j$, $j=1\ldots N_{g}$,
satisfy\begin{eqnarray}
\frac{dM_{i}^{(\alpha)}}{dt} & = & -M_{i}^{(\alpha)}+M_{sp}+S_{i}^{(\alpha)}-w\sum_{j=1}^{N_{g}}W_{ij}^{(mg)}\left[G_{j}^{(\alpha)}\right]_{+},\label{eq:M_eq}\\
\frac{dG_{j}^{(\alpha)}}{dt} & = & -G_{j}^{(\alpha)}+\sum_{k=1}^{N_{m}}W_{jk}^{(gm)}\left[M_{k}^{(\alpha)}\right]_{+},\label{eq:G_eq}\end{eqnarray}
with the rectifier defined as \begin{equation}
[x]_{+}=\begin{cases}
0 & \qquad\mbox{for }x<0\\
x & \qquad\mbox{for }x\ge0.\end{cases}\label{eq:rectifier}\end{equation}
Here $M_{sp}$ denotes the spontaneous activity of the mitral cells
\cite{RiKo06a,YaJu07} in the absence of any odor stimulus and without
any inhibitory inputs from granule cells. The stimuli $S_{i}^{(\alpha)}$
are taken from a stimulus ensemble $\mathbf{S}=\{S_{i}^{(\alpha)},\ \alpha=1\ldots N_{s}\}$.
 Throughout this paper  we consider only  the steady states that the
mitral and granule cell activities reach in response to  long stimulus
presentations. The temporal evolution that we discuss is that of the
network connectivity, which has a time scale that is much slower than
that of the neuronal activities. 

For strong inhibition and in particular for asymmetric connectivity
matrices (cf. \textit{RESULTS, Imperfect Reciprocity of Synapses is Sufficient}) 
the steady states of (\ref{eq:M_eq},\ref{eq:G_eq})
may become unstable. While we did find complex eigenvalues in the
spectrum of the linear operator of (\ref{eq:M_eq},\ref{eq:G_eq}),
which correspond to oscillatory modes, in none of the cases we considered
did any of the eigenvalues have positive real parts. Thus, the steady
output patterns remained stable for all parameters values we considered.

For sufficiently large spontaneous activity $M_{sp}$ essentially
all mitral cell activities -- and with them the granule cell activities
-- are positive and the results are changed only slightly if the rectified
coupling Eq.(\ref{eq:rectifier}) is replaced by a linear coupling.
We therefore use in the following only the linear coupling, which
reduces the computational effort substantially.

The restructuring of the network due to neurogenesis is implemented
by adding granule cells to the network at a steady rate $\beta$ and
removing them with an activity-dependent probability. No detailed
information is available to what extent the formation of synapses
between granule and mitral cells depends on their activity or the
previous presence of synapses at that location. Since the secondary
dendrites of the mitral cells, onto which the granule cells synapse,
extend over large portions of the olfactory bulb we assume in this
minimal model that each granule cell has the potential to establish
a connection with any of the mitral cells. Thus, we assume that each
of the new granule cells connects to $n_{connect}$ randomly chosen
mitral cells. Consistent with observations \cite{WoSh91} we assume
that no granule cell connects to any mitral cell twice. 

A characteristic feature of the dendro-dendritic synapses connecting
mitral and granule cells is the prevalent juxtaposition of a
glutamatergic synapse onto the granule cell and a GABAergic synapse
onto the mitral cell \cite{Sh04}. There are indications that the
glutamatergic and the GABAergic component form at very similar times
\cite{WhGr07},  with possibly the glutamatergic component formed
somehwat earlier \cite{PaBa09}.  Anatomically, the connections between
these two cell types are therefore predominantly reciprocal. This
reciprocity is  important for the ability of the resulting network to
decorrelate stimulus representations.  It implies \begin{equation}
W_{ji}^{(mg)}=W_{ij}^{(gm)}\label{eq:reciprocal}\end{equation} with
$W_{ij}^{(gm)}=1$ if granule cell $i$ is receiving input from mitral
cell $j$ and $W_{ij}^{(gm)}=0$ otherwise. Due to the linearity of the
neuronal dynamics the effective connectivity matrix is given by
\begin{equation}
\boldsymbol{\mathcal{W}}=\mathbf{W}^{(mg)}\mathbf{W}^{(gm)}.\label{eq:effective_connectivity}\end{equation}
Focusing on the structural plasticity provided by the persistent
turn-over of granule cells rather than any plasticity of the synapses
\cite{SaHo06,GaSt09}, we assume in most of this work that all
inhibitory synapses have fixed equal strength $w$ and all excitatory
synapses have strength 1. To probe the role of reciprocity we consider
in Sec.~\textit{RESULTS; Imperfect Reciprocity of Synapses is
Sufficient} also connectivities violating (\ref{eq:reciprocal}) and
heterogeneities in the synaptic strength $w$. 

We model activity-dependent apoptosis of the granule cells \cite{LiSi10}
by discrete events \cite{YokYam11} during which the survival of any
given granule cell is assessed based on the history of its activity.
The duration of the time interval over which the activity influences
cell survival is currently not known. We assume that it is long enough
for the animal to be exposed to a number of relevant odors defining
a stimulus ensemble \textbf{S}. Thus, at each of these events, which
we assume for simplicity to occur regularly in time defining  a time
step of length $\Delta T=1$, granule cells are removed probabilistically.
 Their survival probability $p$ is taken to depend in a sigmoidal
fashion on their cumulative, thresholded activity across the stimulus
ensemble $\mathbf{S}$. Introducing the resilience $R_{i}$ of granule
cell $i$ via\begin{equation}
R_{i}=\sum_{\alpha=1}^{N_{s}}\left[G_{i}^{(\alpha)}-G_{min}\right]_{+}\label{eq:resilience}\end{equation}
with a resilience threshold $G_{min}$, we take \begin{equation}
p(R_{i})=p_{min}+\frac{1}{2}\left\{ \tanh\left(\gamma\left(R_{i}-R_{0}\right)\right)+1\right\} \left(p_{max}-p_{min}\right)\label{eq:survival}\end{equation}
with a soft survival threshold $R_{0}$. Since little is known about
the specifics of the survival probability we take here $p_{min}=0$
and $p_{max}=1$. In this model a granule cell has to reach an activity
beyond $G_{min}$ at least for some of the stimuli in the ensemble
in order to trigger the signaling pathway that controls its survival
\cite{DoPa01,MiSt05,LiSi10}. 

The probabilistic network evolution eventually leads to a statistically
steady state as characterized by the output patterns and their correlations
fluctuating around constant values. The magnitude of the fluctuations
decreases with an increase in the overall number of granule cells
in the system, which can be achieved by a suitable decrease in the
synaptic weight $w$. Fig.\ref{fig:enrichment}B shows the typical
size of fluctuations in the correlation for the parameters used in
our study. 

The network evolution is self-regulated by the balance between proliferation
and apoptosis: with increasing granule-cell population the overall
inhibition of the mitral cells increases, leading to a reduction in
granule cell activity. This lowers the survival probability of the
granule cells and provides the saturation of the network size. As
observed experimentally, reduced odor stimulation leads to a reduction
in the size of the granule cell population \cite{PeAl02}.

We quantify the reshaping of the stimulus representations by the resulting
network using the Pearson correlation $r_{\alpha\beta}$ of patterns
$\mathbf{M}^{(\alpha,\beta)}$\begin{equation}
r_{\alpha\beta}=\frac{\sum_{i=1}^{N_{m}}\left(M_{i}^{(\alpha)}-\left\langle \mathbf{M}^{(\alpha)}\right\rangle \right)\left(M_{i}^{(\beta)}-\left\langle \mathbf{M}^{(\beta)}\right\rangle \right)}{\sqrt{\sum_{i=1}^{N_{m}}\left(M_{i}^{(\alpha)}-\left\langle \mathbf{M}^{(\alpha)}\right\rangle \right)^{2}}\sqrt{\sum_{i=1}^{N_{m}}\left(M_{i}^{(\beta)}-\left\langle \mathbf{M}^{(\beta)}\right\rangle \right)^{2}}}\label{eq:Pearson}\end{equation}
where $\left\langle \mathbf{M}^{(\alpha)}\right\rangle =N_{m}^{-1}\sum_{i=1}^{N_{m}}M_{i}^{(\alpha)}$.
The average correlation $\bar{r}$ is given by \begin{equation}
\bar{r}=\frac{1}{N_{s}\left(N_{s}-1\right)}\sum_{\alpha\ne\beta}^{N_{s}}r_{\alpha\beta}.\label{eq:mean _pearson}\end{equation}

\subsection*{Population Description}

To capture certain aspects of the network evolution analytically we
also consider the weak-coupling limit, $w\ll1$. The number of granule
cells is then large and the network restructuring can be described
in terms of differential equations for the mean size of the various
populations of granule cells that have established the same connections
with mitral cells. For simplicity we give here only the equations
for two-connection networks in which each granule cell makes connections
with two mitral cells. The probability $P(n_{ij},t)$ for the population
of granule cells connecting mitral cells $i$ and $j$ to have size
$n_{ij}$ evolves during a small time step $\Delta t$ according to
\begin{eqnarray}
P(n_{ij},t+\Delta t) & = & P(n_{ij},t)+\nonumber \\
 &  & \Delta t\left\{ \beta\left[P(n_{ij}-1,t)-P(n_{ij},t)\right]-\delta(n_{ij})P(n_{ij},t)+\delta(n_{ij}+1)P(n_{ij}+1,t)\right\} ,\label{eq:Markov}\end{eqnarray}
where $\beta$ is the fixed influx of new granule cells and $\delta(n_{ij})$
is the removal rate. With $p(R_{ij})$ giving the probability for
a granule cell to survive for the duration $\Delta T=1$, the removal
rate is given by \begin{equation}
\delta(n_{ij})=\frac{1}{\Delta t}n_{ij}\left\{ 1-\left[p(R_{ij})\right]^{\frac{\Delta t}{\Delta T}}\right\} \rightarrow-n_{ij}\ln p(R_{ij})\qquad\mbox{for }\Delta t\rightarrow0.\label{eq:death_rate}\end{equation}
Here we have used that different cells are removed independently of
each other. The resilience $R_{ij}$ is given in terms of the activity
of the granule cells $G_{ij}^{(\alpha)}$ analogous to Eq.(\ref{eq:resilience}).
For large mean population size $\langle n_{ij}\rangle$ the probability
distribution $P(n_{ij}$,t) will be sufficiently sharply peaked to
allow to approximate the evolution equation for  $\langle n_{ij}\rangle$,\begin{equation}
\frac{d\left\langle n_{ij}\right\rangle }{dt}=\beta-\sum_{n_{ij}=1}^{\infty}\delta(n_{ij})P(n_{ij},t),\label{eq:pop_model_1}\end{equation}
by \begin{equation}
\frac{dn_{ij}}{dt}=\beta+n_{ij}\ln p\left(R_{ij}\right).\label{eq:population_model}\end{equation}
Here and in the following we drop the brackets indicating the mean
value. 

The steady-state neuronal activities are given by \begin{eqnarray}
M_{i}^{(\alpha)} & = & M_{sp}+S_{i}^{(\alpha)}-w\sum_{i\ne j=1}^{N_{m}}n_{ij}G_{ij}^{(\alpha)}\label{eq:population_MC}\\
G_{ij}^{(\alpha)} & = & M_{i}^{(\alpha)}+M_{j}^{(\alpha)}.\label{eq:population_GC}\end{eqnarray}

Note that for networks with a realistic number of mitral cells the
number of possible different granule cell populations is extremely
large, much larger than the total number of granule cells in the olfactory
bulb. Thus, the size of most populations will be small and fluctuations
in the number of granule cells, which have been neglected in the population
description eqs.(\ref{eq:population_model},\ref{eq:population_MC},\ref{eq:population_GC}),
may become relevant. The main purpose of the population formulation
is to allow analytical approaches for simple cases, which can provide
insight that may be hard to extract from numerical simulations of
the discrete model. When interpreting the analytical results the limitations
of the formulation need to be kept in mind.  

For analytical calculations considering a steep sigmoid, $\gamma\gg1$,
for the survival probability $p(R_{i})$ in Eq.(\ref{eq:population_model})
is particularly attractive. The steady state of the population description
eqs.(\ref{eq:population_model},\ref{eq:population_MC},\ref{eq:population_GC})
can then be analyzed quite easily because the nullcline for the population
$n_{ij}$, which is defined by $\frac{dn_{ij}}{dt}=0$, is then very
well approximated by $R_{ij}=R_{0}$ since $\ln\left(p(R_{ij})\right)$
switches quickly from $0$ to $-\infty$ as $R_{ij}$ passes through
$R_{0}$.

\subsection*{Interference and Optimal Resilience Threshold}

To obtain analytical results for the threshold $G_{min}^{(opt)}$
that minimizes interfering connections between mitral cells that are
strongly active but only during the presentation of different stimuli
we consider a set of four stimuli \textbf{$\mathbf{S}^{(\alpha)}$}
activating four glomeruli,\begin{eqnarray}
\mathbf{S}^{(1,2)} & = & \left(S\pm s,S\mp s,0,0\right),\label{eq:4_stimuli_1}\\
\mathbf{S}^{(3,4)} & = & \left(0,0,S\pm s,S\mp s\right).\label{eq:4_stimuli_2}\end{eqnarray}
The symmetry of this stimulus ensemble has been chosen such that for
networks in which each granule cell connects to two mitral cells only
two granule-cell populations have to be analyzed, $n_{12}$ and $n_{13}$.
Independent of the values of the thresholds the remaining populations
are given by \[
n_{34}=n_{12}\qquad\mbox{and }\qquad n_{14}=n_{23}=n_{24}=n_{13}.\]
For $s\ll S$ these stimulus pairs are highly correlated. We consider
their reshaping by networks that are trained using the slightly simplified
ensemble $\{\mathbf{S}^{(i)},i=1\ldots4\}$ with $s=0$. The approximate
 nullclines $R_{12}=R_{0}$ and $R_{13}=R_{0}$ for the evolution
of the two populations $n_{12}$ and $n_{13}$ are then given by (cf.
Eq.(\ref{eq:resilience})) \begin{align}
R_{0} & =2\left[\frac{2}{2n_{12}+4n_{13}+1}\left(\frac{2n_{12}+1+2n_{13}}{2n_{12}+1}S+M_{sp}\right)-G_{min}\right]_{+}\nonumber \\
 & +2\left[\frac{2}{2n_{12}+4n_{13}+1}\left(-\frac{2n_{13}}{2n_{12}+1}S+M_{sp}\right)-G_{min}\right]_{+}\label{eq:R12}\\
R_{0} & =4\left[\frac{S+2M_{sp}}{2n_{12}+4n_{13}+1}-G_{min}\right]_{+}.\label{eq:R13}\end{align}
Without loss of generality we have absorbed $w$ into the definition
of $n_{ij}$. Depending on $G_{min}$ the system has two fixed points.
For $G_{min}<$$G_{min}^{(opt)}$ one has \begin{equation}
n_{12}^{(1)}=\frac{4S-R_{0}}{2R_{0}}\qquad n_{13}^{(1)}=4S\frac{G_{min}^{(opt)}-G_{min}}{R_{0}\left(4G_{min}+R_{0}\right)},\label{eq:fixed_point_population}\end{equation}
where $G_{min}^{(opt)}$ is given by \textbf{\begin{equation}
G_{min}^{(opt)}=\frac{1}{2}\frac{R_{0}M_{sp}}{S}.\label{eq:optimal_Gmin-1}\end{equation}
}For $G_{min}>G_{min}^{(opt)}$ the fixed point is given by \begin{equation}
n_{12}^{(2)}=\frac{2(S+M_{sp})}{2G_{min}+R_{0}}-\frac{1}{2}\qquad n_{13}^{(2)}=0.\label{eq:fixed_point_n13_vanishes}\end{equation}
 Thus, the interference induced by population $n_{13}$ vanishes for
$G_{min}>G_{min}^{(opt0}$, while the inhibition of the co-active
cells starts to decrease at $G_{min}=G_{min}^{(opt)}.$ Since the
correlation $r$\textbf{ }decreases with decreasing $n_{13}$ but
increases with decreasing $n_{12}$ it is minimal for $G_{min}^{(opt)}$.

Two comments regarding the solution (\ref{eq:fixed_point_n13_vanishes})
with $n_{13}=0$ are in order. The nullclines are given by (\ref{eq:R12},\ref{eq:R13})
only in the limit $\gamma\rightarrow\infty.$ For finite values of
$\gamma$ corrections arise that render $n_{13}$ non-zero (cf. (\ref{eq:population_model})).
Moreover, the description of the granule cell populations solely in
terms of their mean values requires that the means are sufficiently
large. In particular, since the population is always non-negative
its mean cannot strictly vanish. Nevertheless, for small influx $\beta$
and large $\gamma$ the population $n_{13}$ will become very small
as $G_{min}$ is increased beyond $G_{min}^{(opt)}$.

\subsection*{Alternating vs Mixture Protocol}

A simple model with four stimuli and four glomeruli can also be used
to obtain insight into the difference between the alternating stimulus
protocol and the mixture protocol of Sec.~\textit{RESULTS; Effective Enrichment: Overall Overlap is Not Sufficient}.
To mimic the alternating protocol we use stimuli (\ref{eq:4_stimuli_1},\ref{eq:4_stimuli_2})
with $s=0$ and for the training with the mixture protocol we use
$\mathbf{S}^{(1,2,3,4)}=(\frac{1}{2}S,\frac{1}{2}S,\frac{1}{2}S,\frac{1}{2}S)$.
For this protocol all granule cell populations are equal, which we
denote by $n_{m}$.

The nullcline determining $n_{m}$ is given by \begin{equation}
R_{0}=4\left[\frac{S+2M_{sp}}{1+6n_{m}}-G_{min}\right]_{+}.\label{eq:nullcline_nm}\end{equation}
For $G_{min}<G_{min}^{(opt)}$ comparison with $R_{13}$ in the alternating
protocol (\ref{eq:R13}) gives $3n_{m}=n_{12}+2n_{13}$, implying 
\begin{equation}
n_{13}^{(1)}<n_{m}<n_{12}^{(1)},\label{eq:mixture_vs_alternating_population}\end{equation}
where $n_{12}^{(1)}$ and $n_{13}^{(1)}$ are the granule cell populations
given by (\ref{eq:fixed_point_population}). 
Thus, within this simple model the mixture protocol induces stronger
inhibition than the alternating protocol between the first and second
pair of mitral cells, $n_{m}>n_{13}^{(1)}$. This inhibition enhances
the decorrelation of stimuli like $(\frac{1}{2}S+s,\frac{1}{2}S+s,\frac{1}{2}S-s,\frac{1}{2}S-s)$
and $(\frac{1}{2}S-s,\frac{1}{2}S-s,\frac{1}{2}S+s,\frac{1}{2}S+s)$,
which mimic the test stimuli of Sec.~\textit{RESULT; Effective Enrichment: Overall Overlap is Not Sufficient}.
This relationship among the populations is also found in the simulations
of the full discrete model of Fig.\ref{fig:protocol}. Excluding the
terms on the diagonal, which provide self-inhibition, the sums of
the synaptic weights in the four quadrants of the effective connectivity
matrix $\boldsymbol{\mathcal{W}}$ are found to be \begin{equation}
\hat{\boldsymbol{\mathcal{W}}}_{alt}=\left(\begin{array}{cc}
\overbrace{145000}^{\backsimeq n_{12}} & \overbrace{101000}^{\backsimeq n_{13}}\\
101000 & 146000\end{array}\right)\quad\hat{\boldsymbol{\mathcal{W}}}_{mix}=\left(\begin{array}{cc}
\overbrace{124000}^{\backsimeq n_{m}} & \overbrace{119000}^{\backsimeq n_{m}}\\
119000 & 120000\end{array}\right).\label{eq:blocks_of_populations}\end{equation}
The terms above the braces indicate which population is considered
to correspond to which block. Note that each quadrant contains many
connections that are not specific to limonene or carvone (cf. Fig.\ref{fig:enrichment});
neither will they be affected much by the difference in the protocol
nor will they contribute substantially to the discrimination of the
test stimuli.

\subsection*{Natural Stimuli }

To test the ability of the model network to decorrelate stimulus representations
we use an ensemble of stimuli modeled after the activation patterns
in the glomerular layer of rat that have been obtained experimentally
via $[^{14}C]2$-deoxyglucose uptake in response to odor exposure
(published in the Glomerular Activity Response Archive http://gara.bio.uci.edu/,
cf. \cite{JoLe07}). In these data the individual glomeruli have not
been identified. Clearly, not each of the\textbf{ $357\times197$
}pixels corresponds to a glomerulus. We have down-sampled the experimentally
determined pixel patterns to 424\textbf{ }input channels (or 50 channels
in cases in which we illustrate the connectivity), and take each channel
as a proxy for a glomerulus. In the down-sampling we avoid excessive
smoothing of the resulting patterns by retaining in each set of adjacent\textbf{
$10\times10$} pixels the highest value rather than their average
(Fig.\ref{fig:mouse_stimuli_decorrelation}A). The stimulus set $\mathbf{S}$
includes 2 pairs of enantiomers, $\pm$-limonene and $\pm$-carvone,
which are difficult to discriminate. Specifically, without training
mice do not discriminate between the two enantiomers of limonene \cite{MoLi09}.
When addressing the ability of the model network to learn to decorrelate
highly similar stimuli we focus on these 4 stimuli. In addition, to
mimic a background odor environment we include four additional stimuli,
1-butanol, 1-hexanol, 1-heptanol,  and acetic acid. 

% Do NOT remove this, even if you are not including acknowledgments
\section*{Acknowledgments}
% HR gratefully acknowledges support by the Alexander-von-Humboldt foundation,
% NIH (1F33DC8064-1), and NSF (DMS-0719944). %
We thank W.L. Kath and R.W. Friedrich for feed-back on the manuscript.

% \section*{References}
% The bibtex filename
\bibliography{journal}

\newpage
 
\section*{Supplementary Information: Comparison of Different Types of Adaptive Networks}
 
Various types of adaptive networks have been discussed in the context
of stimulus discrimination by the olfactory bulb. In an early model
for the restructuring of the inhibitory bulbar network by neurogenesis
granule cells were not treated explicitly. Instead, inhibition was
assumed to be pairwise and symmetric between mitral cells with a strength
that was taken as a proxy for the number of granule cells connecting
the two mitral cells \cite{CePe01}. The mitral cells were described
with a linear firing-rate model. The inhibitory weight of the connection
between two mitral cells was increased proportional to the pairwise
scalar product of the activity of these two mitral cells across a
stimulus ensemble. Since the activities represented deviations from
the mean they could also be negative in this model. 

From a more general perspective it has been shown that a network can
make stimulus representations orthogonal if it normalizes the activity
of each output channel (i.e. the activity of each mitral cell) across
the stimulus ensemble and orthogonalizes the output activities \cite{WiWi10}.
In a recurrent inhibitory network with symmetric connectivity this
is achieved if the inhibition between the channels is essentially
given by the square-root of the matrix of pairwise scalar products
of the input activities. 

For the honey bee olfactory system experimental data are available
for the inputs as well as the outputs of the antennal lobe. These
were employed in a computational model of the antennal lobe to show
that the pattern transformation performed by the antennal lobe is
better captured by models in which the connectivity of the inhibitory
network is based on the correlations between the glomerular inputs
than by models with random or spatially local connectivity \cite{LiSa05}. 

Motivated by these different types of adaptive networks we compare
the ability of a number of different adaptive recurrent networks to
decorrelate the natural stimuli of Fig.2.
In all cases only the principal neurons are retained and their dynamics
are given by the linear rate model\begin{equation}
\frac{dM_{i}^{(\alpha)}}{dt}=-M_{i}^{(\alpha)}+M_{sp}+S_{i}^{(\alpha)}-w\sum_{j=1}^{N_{m}}\mathcal{W}_{ij}M_{j}^{(\alpha)}\label{eq:M_comparison}\end{equation}
with an effective inhibitory connectivity matrix $\boldsymbol{\mathcal{W}}$.
We use $\mathbf{S}=\{S_{i}^{(\alpha)},\ \alpha=1\ldots N_{s}\}$ to
denote the ensemble of $N_{s}$ stimuli in terms of a matrix in which
the columns consist of the stimulus activities. In our neurogenetic
model the spontaneous mitral cell activity $M_{sp}$ modifies the
connectivity. In the other connectivities we incorporate $M_{sp}$
by using an effective stimulus ensemble $\hat{\mathbf{S}}=\mathbf{S}+\mathbf{M}_{sp}$,
where $\mathbf{M}_{sp}$ is a matrix in which all entries are equal
to $M_{sp}$. The connectivity matrices are then given by \begin{eqnarray}
\mathcal{W}_{ij}^{(ng)} & = & \mathbf{W}^{(mg)}\mathbf{W}^{(gm)}\qquad\mbox{Neurogenesis}\label{eq:conn_neurogenesis}\\
\boldsymbol{\mathcal{W}}^{(ortho)} & = & \left[\frac{1}{\Gamma}\left(\hat{\mathbf{S}}\mathbf{\hat{S}}^{t}\right)^{\frac{1}{2}}-\mathbf{I}\right]_{+}\qquad\mbox{Orthogonalizing}\label{eq:conn_ortho}\\
\boldsymbol{\mathcal{W}}{}^{(L_{2})} & = & \mathbf{\hat{S}}\mathbf{\hat{S}}^{t}\qquad\mbox{L}_{2}\label{eq:conn_L2}\\
W_{ij}^{(dist)} & = & \frac{1}{\sqrt{\sum_{\alpha=1}^{N_{s}}\left(S_{i}^{(\alpha)}-S_{j}^{(\alpha)}\right)^{2}}}\qquad\mbox{Distance}\label{eq:conn_distance}\\
W_{ij}^{(corr)} & = & \left[r(S_{i},S_{j})\right]_{+}\qquad\mbox{Pearson correlation}\label{eq:conn_pearson}\\
W_{ij}^{(cov)} & = & \left[\mbox{cov}(S_{i},S_{j})\right]_{+}\qquad\mbox{Covariance}.\label{eq:conn_covariance}\end{eqnarray}
In each of these connectivities the inhibition between pairs of mitral
cells is related to a broadly defined similarity of their responses
to the stimuli in the ensemble. The similarity is, however, assessed
in different ways in each case. Connectivity $\boldsymbol{\mathcal{W}}^{(ng)}$
 is obtained from our neurogenesis model 
 (eqs. (3,4,9) in the main manuscript)
 %(\ref{eq:M_eq},\ref{eq:G_eq},\ref{eq:survival})
with 50 mitral cells. In this model the similarity is measured in
terms of an additive co-activity. In connectivities $\boldsymbol{\mathbf{\mathcal{W}}}^{(ortho)}$
and $\boldsymbol{\mathcal{W}}^{(L_{2})}$ the similarity is measured
in terms of a multiplicative co-activity. The connectivity $\boldsymbol{\mathcal{W}}^{(ortho)}$
stems from the orthogonalizing networks \cite{WiWi10}, with the parameter
$\Gamma$ determining the $L_{2}$-norm of the outputs with $w=1$
fixed. Without any attempt to optimize the performance we choose here
$\Gamma=1$. Connectivity $\boldsymbol{\mathcal{W}}^{(L_{2})}$ is
motivated by the algorithm used in \cite{CePe01}. Connectivity $\boldsymbol{\mathcal{W}}^{(corr)}$
is motivated by the modeling of the antennal lobe network \cite{LiSa05};
here $r(S_{i},S_{j})$ denotes the Pearson correlation between the
stimulus activities of glomerulus $i$ and glomerulus $j$ given the
stimulus ensemble $\mathbf{S}$. Analogously, in connectivity $\boldsymbol{\mathcal{W}}^{(cov)}$
$\mbox{cov}(S_{i},S_{j})$ denotes the corresponding covariance. Writing
$\mathbf{S}$ instead of $\mathbf{\hat{S}}$ in $\boldsymbol{\mathcal{W}}^{(corr)}$,
$\boldsymbol{\mathcal{W}}^{(cov)}$, and $\boldsymbol{\mathcal{W}}^{(dist)}$
emphasizes the fact that these connectivities are not sensitive to
$M_{sp}$. The connectivity matrices $\boldsymbol{\mathbf{\mathcal{W}}}^{(ortho)}$,
$\boldsymbol{\mathcal{W}}^{(corr)}$, and $\boldsymbol{\mathcal{W}}^{(cov)}$
are not guaranteed to have only positive entries. We therefore set
any negative entries to 0. 

To compare the different connectivities we use two different approaches.
Since the decorrelation performance often improves with increasing
inhibition and decreasing output amplitude, we choose the overall
inhibitory weight $w$ for each connectivity such that it generates
the same mean output amplitude as the neurogenesis model. Some connectivities
perform, however, optimally at an intermediate inhibitory strength.
Therefore we also consider the dependence of the performance on the
inhibitory strength.

We characterize the different networks in two ways. We determine the
similarity of the connectivities directly using a scaled distance
$\mathcal{D}$ between them, which we define as \begin{equation}
\mathcal{D}(\mathbf{W}^{(1)},\mathbf{W}^{(2)})^{2}=\frac{\sum_{ij}\left(W_{ij}^{(1)}-W_{ij}^{(2)}\right)^{2}}{\frac{1}{2}\left(\sqrt{\sum_{ij}W_{ij}^{(1)2}}+\sqrt{\sum_{ij}W_{ij}^{(2)2}}\right)^{2}}.\label{eq:distance_connectivities}\end{equation}
Then we assess the performance of the networks in terms of the correlations
of their outputs given the stimulus ensemble $\mathbf{S}$. 

The distance measure $\mathcal{D}$ reveals that the co-activity based
connectivities $\boldsymbol{\mathcal{W}}^{(ng)}$, $\boldsymbol{\mathbf{\mathcal{W}}}^{(ortho)}$,
and $\boldsymbol{\mathcal{W}}^{(L_{2})}$ are quite similar to each
other for both values of $M_{sp}$ (Fig.S\ref{fig:distance_connectivities}).
The relationship among the other connectivities is not as clear. For
$M_{sp}=0$ it appears as if $\boldsymbol{\mathcal{W}}^{(dist)}$
and $\boldsymbol{\mathcal{W}}^{(corr)}$ also formed a cluster. However,
it does not persist for $M_{sp}=1$ (Fig.S\ref{fig:distance_connectivities}B)
and other values of the mean output amplitude (not shown). Similarly,
for some output amplitudes $\boldsymbol{\mathcal{W}}^{(corr)}$ and
$\boldsymbol{\mathcal{W}}^{(cov)}$ are much closer to each other
than for the output amplitudes used in Fig.S\ref{fig:distance_connectivities}.

\begin{figure}[!ht]
\begin{center}
\includegraphics[width=14cm]{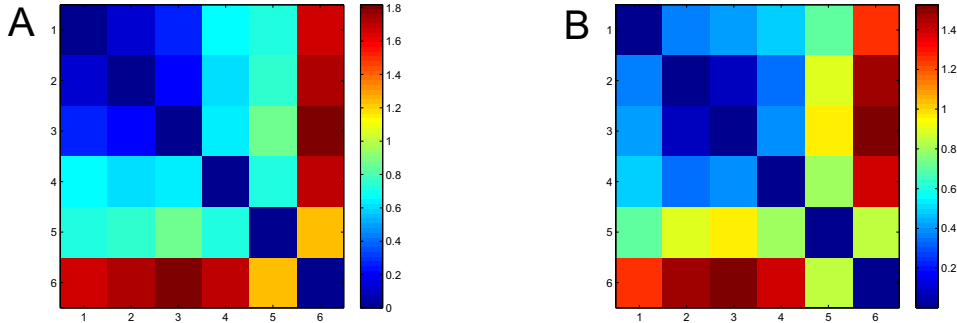}
\end{center}
\caption{
{\bf Similarity of connectivities.} Matrix of scaled pairwise distance $\mathcal{D}$ between the connectivities
(\ref{eq:conn_neurogenesis}-\ref{eq:conn_covariance}) for $M_{sp}=0$
with coupling strengths $\mathbf{w}=(0.002,1.35,0.32,0.13,0.85,70)$
(\textbf{A}) and $M_{sp}=1$ with coupling strengths $\mathbf{w}=(0.002,0.31,0.012,0.105,0.8,50)$
(\textbf{B}). In the distance matrix and the list of coupling strengths
$\mathbf{w}$ the order of the connectivities is $\boldsymbol{\mathcal{W}}^{(ng)}$,
$\boldsymbol{\mathbf{\mathcal{W}}}^{(ortho)}$, $\boldsymbol{\mathcal{W}}^{(L_{2})}$,
$\boldsymbol{\mathcal{W}}^{(dist)}$, $\boldsymbol{\mathcal{W}}^{(corr)},$
$\boldsymbol{\mathcal{W}}^{(cov)}$. With these synaptic weights $w$
the mean output amplitude is the same for all connectivities, $\bar{\bar{M}}\equiv\left(N_{s}N_{m}\right)^{-1}\sum_{i,\alpha}M_{i}^{(\alpha)}=0.0265$\textbf{
}for $M_{sp}=0$ and $\bar{\bar{M}}=0.15$ for $M_{sp}=1$. For $\boldsymbol{\mathcal{W}}^{(cov)}$
and $M_{sp}=1$ the output amplitude had to be chosen larger, $\bar{\bar{M}}=0.189$,
to avoid dynamical instability in (\ref{eq:M_comparison}). Parameters
for the runs generating $\boldsymbol{\mathcal{W}}^{(ng)}$: $N_{m}=50$,
$\gamma=20$, $R_{0}=0.1$, $n_{connect}=4$, $\beta=3.9$, $w=0.002$,
\textbf{A: }$M_{sp}=0$, $G_{min}=0.12$, \textbf{B:} $M_{sp}=1$,
$G_{min}=0.6$. 
}
\label{fig:distance_connectivities}
\end{figure}

To assess the decorrelation performance of the networks directly we
determine the correlations in the outputs for the stimuli to which
they are adapted. Without the spontaneous activity of the mitral cells,
$M_{sp}=0$, each of the connectivities (\ref{eq:conn_neurogenesis}-\ref{eq:conn_covariance})
is able to reduce the correlation of the representations of the highly
similar stimuli (black bars in Fig.S\ref{fig:network_decorrelation_comparison}A),
albeit to very different degrees. This is not the case for the less
correlated stimuli. Consistent with the similarity of the co-activity
based connectivities seen in Fig.S\ref{fig:distance_connectivities},
$\boldsymbol{\mathcal{W}}^{(ng)}$ , $\boldsymbol{\mathbf{\mathcal{W}}}^{(ortho)}$,
and $\boldsymbol{\mathcal{W}}^{(L_{2})}$ perform similarly and quite
well for all stimuli. The connectivities $\boldsymbol{\mathcal{W}}^{(dist)}$,
$\boldsymbol{\mathcal{W}}^{(corr)}$, and $\boldsymbol{\mathcal{W}}^{(cov)}$,
however, only decorrelate the representations of the highly similar
stimuli; for the less related inputs their performance is very poor;
in fact, on average the outputs of $\boldsymbol{\mathcal{W}}^{(corr)}$
and $\boldsymbol{\mathcal{W}}^{(cov)}$ are more correlated than their
inputs (Fig.S\ref{fig:network_decorrelation_comparison}B). 

\begin{figure}[!ht]
\begin{center}
\includegraphics[width=14cm]{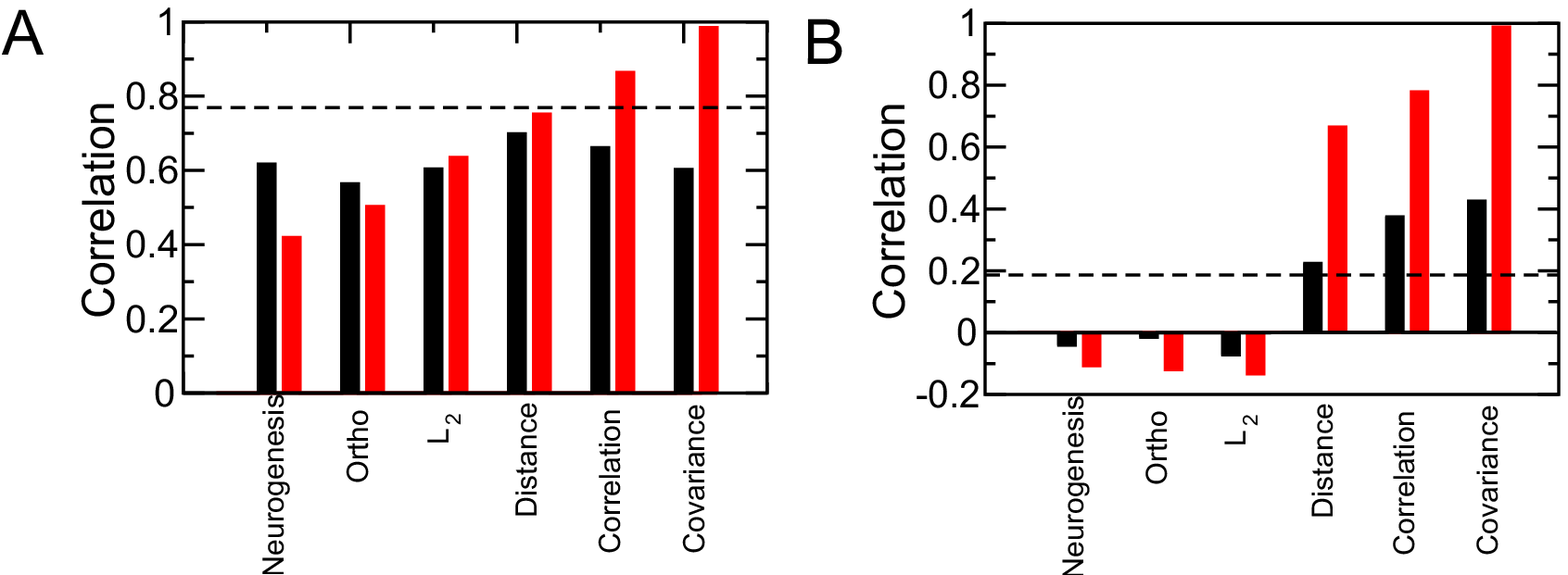}
\includegraphics[width=14.5cm]{suppfig2.eps}
\end{center}
\caption{
{\bf Comparison of the decorrelation performance of adaptive networks.}
\textbf{A) }Top correlations $r^{(top)}$ obtained with the connectivities
(\ref{eq:conn_neurogenesis}-\ref{eq:conn_covariance}). \textbf{B)
}Mean correlations $\bar{r}$ obtained with the connectivities (\ref{eq:conn_neurogenesis}-\ref{eq:conn_covariance}).
$M_{sp}=0$ (black), $M_{sp}=1$(red). Dashed lines denote the corresponding input correlations. 
}
\label{fig:network_decorrelation_comparison}
\end{figure}

For non-zero spontaneous activity, $M_{sp}=1$, most connectivities
perform worse, in particular for the highly similar stimuli. Nevertheless,
the co-activity based connectivities perform still quite well overall.
The performance of $\boldsymbol{\mathcal{W}}^{(corr)}$, $\boldsymbol{\mathcal{W}}^{(cov)}$,
and $\boldsymbol{\mathcal{W}}^{(dist)}$, however, is poor. Even the
representations of the highly similar stimuli are not decorrelated
any more. This reflects, in part, the fact that these adaptation schemes
do not account for the spontaneous mitral cell activity, which effectively
corresponds to a significant mean value in the input activities. 

The strikingly poor decorrelation by the correlation-based and the
covariance-based network for the inhibition level used in Fig.S\ref{fig:network_decorrelation_comparison}
raises the question whether their performance could be improved by
optimizing the overall inhibition. We therefore vary for each network
the overall inhibition level $w$ while keeping the network structures
fixed as defined by (\ref{eq:conn_neurogenesis}-\ref{eq:conn_covariance}).
For adaptive networks the inhibition level $w$ affects, in principle,
also the training of the network; the network structure is therefore
not really independent of $w$. However, none of the comparison networks
(\ref{eq:conn_ortho}-\ref{eq:conn_covariance}) are defined via a
training algorithm; for them $w$ is a free parameter. Varying $w$
over a range that induces comparable changes in the output amplitude
we find that the performance of the co-activity based connectivities
improves monotonically with decreasing amplitude. In Fig.S\ref{fig:corr_weight_dependence}
this is shown for the top correlation $r^{(top)}$ for $M_{sp}=0$
(black) and $M_{sp}=1$ (red). It also holds for the overall correlation
(not shown). The connectivities $\boldsymbol{\mathcal{W}}^{(corr)}$
and $\boldsymbol{\mathcal{W}}^{(cov)}$ exhibit different behavior.
For $M_{sp}=0$ they both decorrelate the representations of the highly
similar stimuli for weak inhibition, but as the inhibition is increased
the output correlation starts to rise again. With non-zero spontaneous
mitral cell activity the output correlation increases monotonically
already starting from $w=0$. We find that the overall correlation
$\bar{r}$ increases monotonically even for $M{}_{sp}=0$ (not shown).
The distance-based connectivity reduces $r^{(top)}$ monotonically
for $M_{sp}=0$. For $M_{sp}=1$, however, the correlation increases
to a maximum before it starts to decrease. Similarly, the overall
correlation $\bar{r}$ increases initially with increasing inhibition. 

\begin{figure}[!ht]
\begin{center}
\includegraphics[width=14cm]{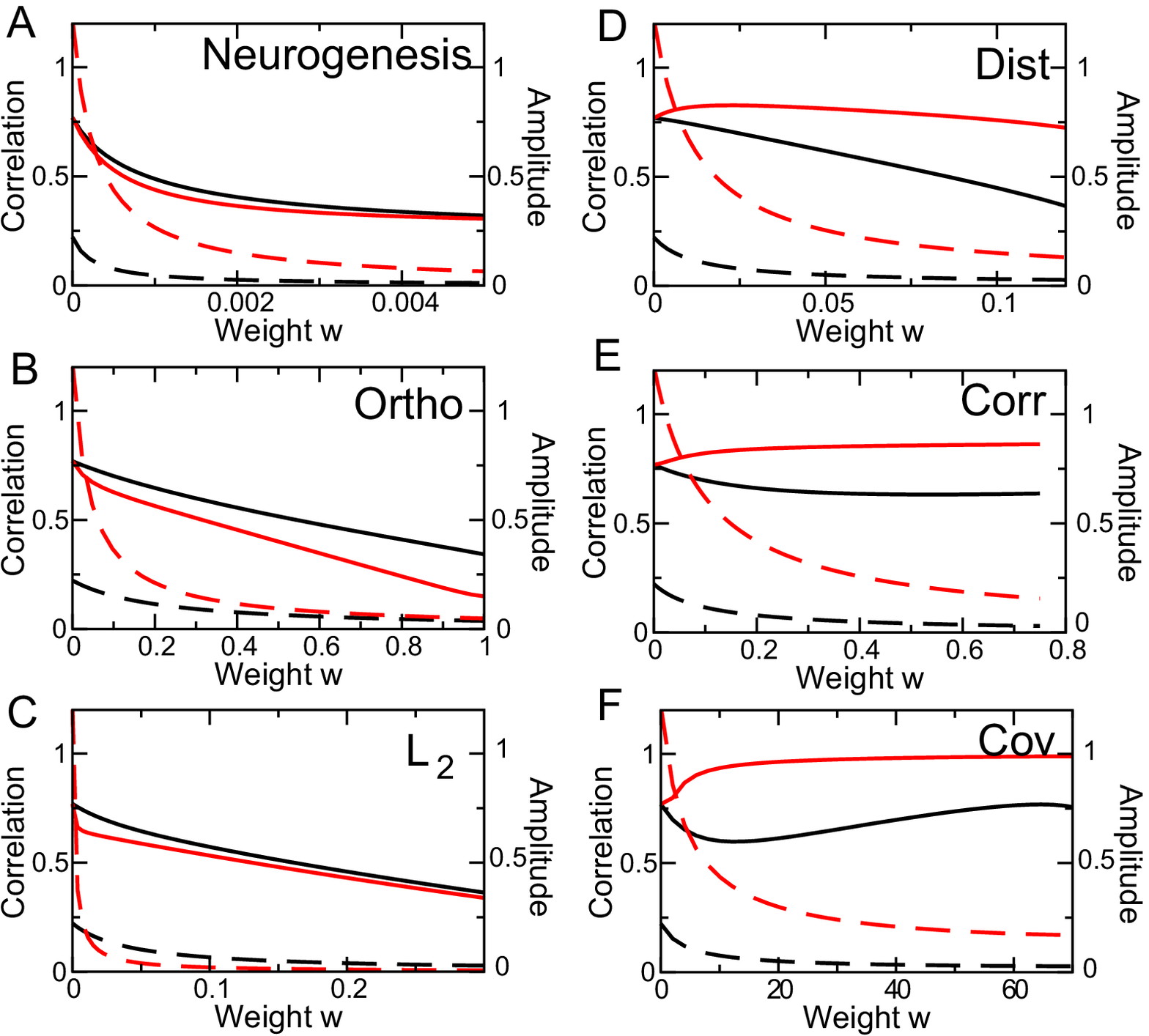}
\end{center}
\caption{
{\bf Dependence of the decorrelation of the highly similar stimuli on the
overall weight $w$.} Black: $M_{sp}=0$, red: $M_{sp}=1$. Solid lines
$r^{(top)}$, dashed lines: mean amplitude. 
}
\label{fig:corr_weight_dependence}
\end{figure}

The network $\boldsymbol{\mathbf{\mathcal{W}}}^{(ortho)}$ is designed
to orthogonalize the representation of the stimulus ensemble. Nevertheless,
in the situations shown in Fig.S\ref{fig:network_decorrelation_comparison}A
and Fig.S\ref{fig:corr_weight_dependence}A it does not achieve this
goal for the highly similar stimuli. This is a result of the requirement
that the network be purely inhibitory, i.e. $W_{ij}^{(ortho)}\ge0$.
Depending on the stimulus ensemble and the desired output amplitudes
this restriction limits its performance to a variable degree. In other
situations essentially perfect decorrelation is obtained \cite{WiWi10}.

Our investigation of connectivities $\boldsymbol{\mathcal{W}}^{(L_{2})}$
and $\boldsymbol{\mathcal{W}}^{(corr)}$ is motivated by previous
work on neurogenesis \cite{CePe01} and odor processing in the antennal
lobe \cite{LiSa05}. It should be noted, however, that these connectivities
do not in detail represent the networks investigated in \cite{CePe01,LiSa05}.
In \cite{CePe01} the effective weights are updated according to the
$L_{2}$ scalar product of the mitral cell activities and additional
self-inhibition as well as a total synaptic weight normalization is
introduced. In \cite{LiSa05} the inhibition is feedforward rather
than recurrent, i.e. it is driven directly by the inputs. 

This survey does not aim to represent an exhaustive analysis of the
various decorrelation mechanisms. It suggests, however, that co-activity
based connectivities like $\boldsymbol{\mathcal{W}}^{(ng)}$ , $\boldsymbol{\mathbf{\mathcal{W}}}^{(ortho)}$,
and $\boldsymbol{\mathcal{W}}^{(L_{2})}$ are much more capable to
decorrelate stimulus representations under various conditions than
correlation- or covariance-based networks.

\end{document}